\documentclass[pdflatex,sn-mathphys-num]{sn-jnl}

\usepackage{lipsum}
\usepackage{graphicx}%
\usepackage{multirow}%
\usepackage{amsmath,amssymb,amsfonts}%
\usepackage{amsthm}%
\usepackage{mathrsfs}%
\usepackage[title]{appendix}%
\usepackage{xcolor}%
\usepackage{textcomp}%
\usepackage{manyfoot}%
\usepackage{booktabs}%
\usepackage{algorithm}%
\usepackage{algorithmicx}%
\usepackage{algpseudocode}%
\usepackage{listings}
\usepackage{enumitem}
\usepackage{subcaption}
\usepackage{caption}
\usepackage{subcaption}
\usepackage{xcolor}

\usepackage{lineno}

\usepackage{geometry}
\newcommand{\figref}[2]{\hyperref[#1]{~\ref*{#1}#2}}

\theoremstyle{thmstyleone}%
\theoremstyle{thmstyletwo}%

\theoremstyle{thmstylethree}%
\newif\ifwithSI
\withSItrue
\ifwithSI\else
  \usepackage{xr}
  \makeatletter
  \newcommand*{\addFileDependency}[1]{\typeout{(#1)}\@addtofilelist{#1}\IfFileExists{#1}{}{\typeout{No file #1.}}}
  \makeatother
  \newcommand*{\myexternaldocument}[1]{\externaldocument{#1}\addFileDependency{#1.tex}\addFileDependency{#1.aux}}
  \myexternaldocument{sn-SI}
\fi

\begin{document}

\title[Article Title]{Efficient Optimization of Mechanical Metamaterials via Transferable Representation Learning}


\author[1]{\fnm{Antonio} \sur{Varagnolo}}\email{avaragnolo3@gatech.edu}
\equalcont{These authors contributed equally to this work.}

\author[2]{\fnm{Miaomiao} \sur{Dai}}\email{mdai49@gatech.edu}
\equalcont{Authors contributed equally}

\author*[2]{\fnm{Bolei} \sur{Deng}}\email{bolei.deng@gatech.edu}

\author*[1]{\fnm{Raphael} \sur{Pestourie}}\email{rpestourie3@gatech.edu}

\affil*[1]{\orgdiv{School of Computational Science and Engineering}, \orgname{Georgia Institute of Technology}, \orgaddress{\city{Atlanta}, \postcode{30332}, \state{GA}, \country{USA}}}

\affil[2]{\orgdiv{Daniel Guggenheim School of Aerospace Engineering}, \orgname{Georgia Institute of Technology}, \orgaddress{\postcode{30332}, \state{GA}, \country{USA}}}


\abstract{Designing mechanical metamaterials for nonlinear behavior such as energy absorption is limited by high-dimensional parametrizations and by expensive simulations. Here we show that design knowledge from a cheap linear property can be transferred to nonlinear, potentially non-differentiable, targets through a learned, reusable search space. TRAMA (TRAnsferable representation learning for Metamaterial Architectures) reparametrizes architected materials using a variational autoencoder with a pre-trained vision transformer and physics-informed augmentations. For energy-absorption efficiency, TRAMA reaches every efficiency level with 95 to 98\% reduction in the evaluations needed by state-of-the-art gradient-free optimization and 3 to 5 fewer than gradient-based optimization. Its best design is comparable to the best found by any baseline with at least 67\% fewer evaluations, which translate into approximately 100 hours saved per campaign with finite-element simulations. Fabricated designs match simulated stress-strain curves within 5.3\%. A post hoc analysis shows where the representation transfers to new properties and where it does not.}

\keywords{Mechanical Metamaterials,  Representation Learning, Inverse Design, Architected Metamaterials}

\maketitle 

\section{Introduction}\label{sec1}


\noindent We engineer a data-driven search space that transfers design knowledge from the linear Young's modulus to nonlinear responses, such as energy absorption and stress-strain curves, accelerating the optimization of properties that are expensive to evaluate and possibly non-differentiable. The latent space effectively reparametrizes an architected material, yielding a beneficial alignment between computationally feasible and fabricable designs that generic
density-based search spaces fail to provide. When maximizing the energy absorption efficiency of a metamaterial, the
proposed method reaches every efficiency on average with $95$ to $98\%$ less evaluations than state-of-the-art
gradient-free method (CMA-ES) and with $66$ to $80\%$ fewer evaluations than a gradient-based method (CCSA). Considering finite-element simulation costs, the method yields
computational savings from the very first design campaign, translating into clock-time savings of about $100$ hours
against the gradient-based alternative and of about $1800$ hours against the gradient-free one, the only option available in
the absence of a differentiable solver. Furthermore, we conduct a post hoc analysis of transferability to new
properties to develop interpretable intuitions about the scope and limitations of these methods.\\

\noindent Mechanical metamaterials derive their effective properties primarily from geometry rather than material composition, creating a powerful opportunity to inverse-design structure instead of synthesizing new materials. Geometry-based design has already enabled advanced functionalities such as multistability~\cite{shan2015multistable}, energy absorption~\cite{zhang2019energy, portela2021supersonic, chen2021reusable}, and programmability~\cite{haghpanah2016multistable, florijn2014programmable, jin2020kirigami, chen2021reprogrammable, medina2020navigating}, with applications in stretchable electronics~\cite{fan2014fractal, ma2017design, herbert2018soft}, tissue engineering scaffolds~\cite{chan2008scaffolding}, and soft robotics~\cite{pal2021exploiting}. With rapid advances in additive manufacturing and fabrication, the central bottleneck has increasingly shifted from fabrication to design~\cite{kumar2020inverse, gu2018tunable, zeng2023inverse, rodriguez2023heterogeneous}. \\
 
\noindent Inverse design of metamaterials remains challenging because geometric design spaces are extremely high-dimensional and structure-property solvers are computationally expensive. This combination makes exhaustive search computationally expensive. When the gradient is readily available, classical tools for inverse design include shape optimization ~\cite{zolesio1992introduction} and topology optimization ~\cite{sigmund2013topology, zeng2023inverse}. Topology optimization designs directly in pixel space, which yields a rich variety of structures but creates a harder trade-off between performance and fabricability. For instance, it often optimizes a continuous density, producing gray-scale designs that must be converted into clean solid shapes before manufacturing, typically at the cost of performance, whereas parametrized inverse design works with manufacturable shapes from the start~\cite{hammond2025unifying}. Moreover, as a local optimization method, it typically requires multiple restarts with different initializations to escape poor local optima. Without gradient access, classical approaches include genetic and evolutionary algorithms~\cite{beyer2002evolution, hansen2003reducing}, which have been successfully applied to metamaterial design~\cite{rodriguez2023heterogeneous, rodriguez2025mechanics}. However, as design variables increase and without further assumptions on the underlying function, gradient-free methods face exponential growth in computational cost, requiring substantially more evaluations to explore the design space due to the curse of dimensionality ~\cite{hastie2009elements}. \\

\noindent Deep Learning has been widely adopted to accelerate the evaluation of the design performance ~\cite{zheng2023deep, zong2026deep} through the use of fast forward structure-property prediction ~\cite{sun2022machine, gu2018bioinspired, ha2023rapid, liang2026machine, deng2022inverse, yang2021deep, zheng2023deep, zhang2021physics, zhang2022efficient} or generative models ~\cite{wang2020deep, zhang2025generative}. However, the acceleration of the exploration of the feasible space needs to amortize the thousands of PDE simulations needed to generate the dataset~\cite{pestourie2025fast}. Scientific Machine Learning (SciML) Physics-informed machine learning (PIML) approaches have attempted to reduce simulation-based design costs by decreasing the number of required physical evaluations or accelerating computations. Key strategies include physics-informed neural networks (PINNs)~\cite{raissi2019physics}, solver correction methods~\cite{mcbane2021component, zanardi2024scalable}, and operator learning ~\cite{li2020fourier, lu2021learning, kovachki2023neural}. However, despite these advances, most frameworks still prioritize forward prediction-mapping designs to properties rather than directly optimizing the design process~\cite{takeishi2021physics}. Moreover, the optimizer can drive the design toward regions where the surrogate is inaccurate, such as those poorly covered by the training data, decreasing the realiability of the method. 

\noindent A further limitation is that deep learning models struggle to transfer knowledge across design tasks or material properties, so optimizing a new property, for instance in a multi-objective framework, requires generating a new dataset, adding to the training cost. To address this, transfer learning ~\cite{zhuang2020comprehensive} has been successful in reducing dataset construction costs in metamaterials~\cite{li2023mechanics, zhang2022predicting, apaza2024transfer, liu2026transfer}, for instance in multi-objective optimization~\cite{apaza2024transfer} or across different polymer composites~\cite{zhang2022predicting}. Other paths forward to reduce the cost of dataset construction in training a surrogate is the use of multi-fidelity ~\cite{you2026multi, lu2022multifidelity} or meta-learning approaches \cite{finn2017model, nichol2018first, varagnolo2026interpretable}. Finally, an emerging direction is concerned with creating re-usable learned representations across diverse properties. This approach has shown promising results in other adjacent fields such as free-space optics ~\cite{ma2019probabilistic, marzban2025hilab, marzban2026inverse, huang2026neural}, and has also been applied in mechanical metamaterials ~\cite{wang2020deep, ye2024predicting}. Ref. ~\cite{ye2024predicting} constructs a latent space using a Variational Autoencoder (VAE) and then feeds the latent features to a deep neural network to predict properties of the geometry. However if applied to learn the physical response, this strategy inherits the concerns that affect other supervised learning strategy, such as reliability out of distribution and transferability to new properties. In contrast we use a representation to discover an efficient search space. Ref.~\cite{marzban2025hilab} learnt a representation of topologies and transfered to similar optical properties. In contrast, our representation reparameterizes architected materials for a transfer between significantly different linear-to-nonlinear properties. Ref.~\cite{wang2020deep} trains a VAE on topologies optimized for stiffness on a 50$\times$50 pixel grid, jointly fitting regressors to the stiffness properties so that the resulting low-dimensional latent space is continuous and clusters microstructures by property, enabling interpolation and a distance metric for shape similarity. However, the representation is used only for the linear elastic properties it was fit on, and is not transferred to a different nonlinear or dynamic property. Moreover, it heavily relies on post-processing tools to ensure fabricability, for example with thresholding and removal of isolated pixels or with homogeneization of non-periodic microstructures into macro-scale designs. In contrast, this work focuses on transferability and its computational benefits for parameterized structures, that align better feasible space and fabricable space.\\

\noindent Here, we introduce TRAMA (TRAnsferable representation learning for Metamaterial Architectures), a linear-to-nonlinear design strategy that uses a dataset of designs, efficiently optimized on cheap linear mechanical properties, to learn a compact and physically-informed representation of the metamaterial design space. We then transfer this representation to computational costly nonlinear inverse design. We demonstrate the approach on rotating-quadrilateral metamaterials, a class of architected materials known for rich nonlinear behavior while still ensuring fabricability ~\cite{deng2022inverse}. Working directly with an architected metamaterial inherently aligns the space of computationally feasible designs with that of fabricable designs. The alignment avoids the postprocessing steps that density-based optimization requires, which improve fabricability but typically degrade performance. First, we perform inverse design on the linear Young’s modulus, using adjoint-based optimization. Because the linear problem is inexpensive and easily differentiable, we efficiently generate a large dataset of geometries spanning a broad spectrum of elastic responses. This dataset serves as a source of knowledge of how geometric features control basic mechanical behavior. We then train a variational autoencoder to learn a low-dimensional latent space of these optimized designs. This latent space serves as a reparameterized design domain that has $40$ times fewer latent variables than the original parameterized space while retaining physically meaningful design variations. We use this learned latent space for nonlinear inverse design, targeting full stress-strain responses under large deformation. Instead of requiring a differentiable nonlinear simulator, which may not be available for more complex nonlinear properties, for instance fracture ~\cite{shaikeea2022toughness}, we perform gradient-free optimization, such as Bayesian optimization, directly in the compact latent space. Across nonlinear design tasks involving rotation, instability, and buckling, our approach substantially outperforms conventional gradient-free optimization in the original design space and achieves performance comparable to, or even exceeding, gradient-based methods when such methods are available. These results suggest that linear mechanical properties may serve as an efficient source of geometrical knowledge for nonlinear metamaterial design, enabling scalable inverse design of complex architected materials without requiring differentiable nonlinear simulation. Pushing this approach to its limit, we provide a post-hoc transferability analysis that informs how linear-to-nonlinear transfer might perform. 

\section{Results}
We illustrate our method on the efficient design of parametrized metamaterials structures. In Sec.~\ref{sec:space}, we showcase the characteristics of the latent space we constructed and we ablate the components that make up our pipeline. Sec.~\ref{sec:energy} reports the successful transfer to the nonlinear energy efficiency maximization problem, while in Sec.~\ref{sec:snap-curve}, we extend to the more challenging nonlinear design that matches desired stress-strain curves. In both sections, the design performance aligns closely with experimental validations. Sec.~\ref{sec:ablation} studies the translation capabilities of our latent space a posteriori on two of design tasks.

\subsection{Creation of the Latent Space}
\label{sec:space}

\begin{figure}
    \centering
    \includegraphics[width=\linewidth]{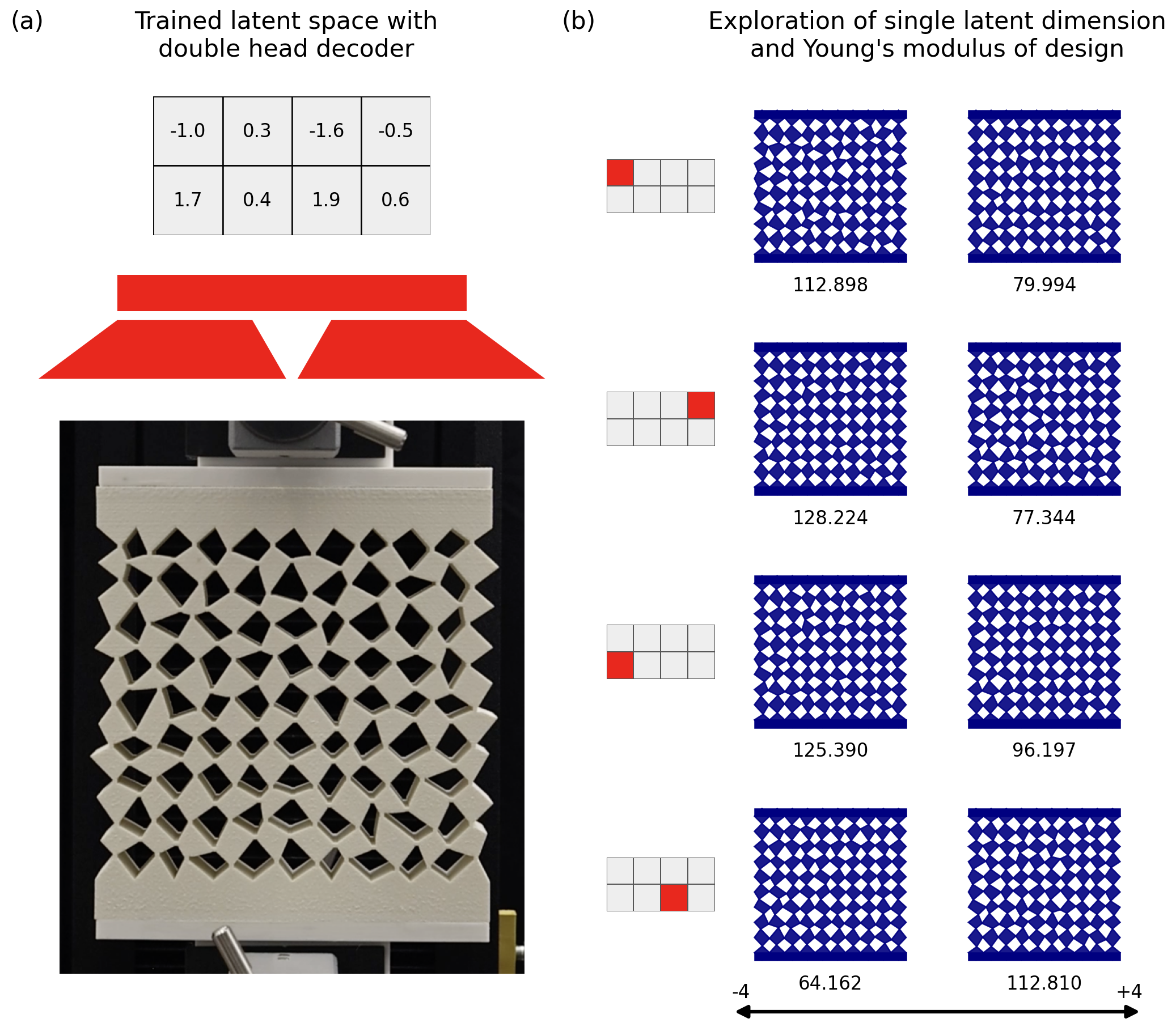}
    \caption{\textbf{TRAMA's decoder and latent space exploration} (a) The two-head decoder takes an 8 dimensional vector and outputs an architected metamaterial. Here we showcase the fabricated version of one of the designs. (b) Exploration of a single latent dimension, and young modulus of each observed structure.}
    \label{fig:latent}
\end{figure}

A diverse initial design library was generated by performing approximately 100 short inverse‑design optimization (40 iterations each, well-before convergence) for the metamaterial's Young's modulus. The Young modulus of the metamaterial is a linear, easily differentiable and inexpensive property to compute. We deploy a linearized version of the solver that outputs the young modulus and its gradient with respect to the design parameters in 1.6 ms (see Supplementary Note~\ref{sec:linearized-modulus} for the detailed implementation). The choice of Young Modulus seemed natural in this application as a property providing inductive bias given its simplicity and widespread applicability in mechanics. Different gradient‑based optimizers were employed, with varying Young's modulus targets (minimization, maximization, and linearly interpolated values). In particular, the three optimizers that we used were: Adam with an augmented Lagrangian treatment of hard constraints ~\cite{AUGLAG}, a Conservative Convex Separable Approximations optimizer (CCSA)~\cite{CCSA}, specifically with surrogates as in the Method of Moving Asymptotes, and SLSQP ~\cite{SLSQP}. For the two latter algorithms we use the nonlinear optimization library \texttt{nlopt} \cite{NLopt}. To ensure fabricability, hard constraints on nodal displacements are applied. Specifically, we enforce hard constraints to avoid node crossing, and on rotating quadrilaterals minimum side sizes as well as minimum angle formed by adjacent quadrilaterals sides (see the Supplementary Note~\ref{sec:design-space} for details). The design library was then enriched through physically informed augmentations. We included mirror symmetries, uniform translations of all nodes, and the addition of alternating perturbations mimicking periodic modes (see Fig.~\ref{fig:augm}), as well as combinations of the above. Translations and alternating perturbations were applied at six magnitudes,
$\{0.01, 0.02, 0.03, 0.05, 0.075, 0.10\}$ in units of the cell size. Larger magnitudes were not used because they almost always violate the node-displacement constraint. Each augmented design is built from a chain of one, two or three operations, applied in random order,
that always belong to different types. A chain length of one, two or three was drawn with probability $0.15$, $0.45$ and $0.40$, respectively.
A candidate was kept only if it satisfied the same node-displacement constraint as the original designs and
was at least $0.05$ away (Euclidean distance in parameter space) from its source design and from all variants already kept. As showcased in Fig.~\ref{fig:augm}c, the Young's modulus along the vertical direction is unaffected by symmetries with vertical axis, and it is enhanced by the addition of alternating perturbations that favor rotation. The transformations allowed us to go from $134$ optimization runs to $\sim 30000$ final training instances (a $\times 235$ increase). 
The ViT-VAE was trained on these designs to compress the 324-dimensional geometry space (144 horizontal and 180 vertical displacement parameters) into an 8-dimensional latent representation, a roughly 40-fold reduction in dimension. The ViT that the encoder was pre-trained on ImageNet-21k and fine-tuned on ImageNet-1k~\cite{dosovitskiy2021image}; its 196 patch tokens are pooled by a learned attention into a 768-dimensional feature vector before the latent projection. The decoder is split into two separate MLP heads (two hidden layers of 1024 units each). Using a pre-trained encoder reduces the data requirement, at the cost of slightly slower training: at the beginning, a transformer trained for generic image tasks struggles to distinguish the design images, and first optimizes only the mean of the designs. Only in later epochs does the model start to fit more refined features. A scheduled KL-divergence weight balances reconstruction fidelity and latent space structure. Fig.~\ref{fig:latent} shows the effect of traversing selected axes of the 8-dimensional latent space. On GPU, the training of the VAE takes around 20 minutes. More information on the training is provided in the Supplementary Note~\ref{sec:exp-appendix}. We evaluate the usefulness of the augmentations and of the pre-trained visual transformer model by comparing the reconstruction ability on a hold-out set. We split the original designs into 3 data folds, and assign each variant to the same fold, so no variants of the same original design are split between training and test set. We first compare the reconstruction performance with and without augmentations on the original, unaugmented, designs. We observe that in 3 out of 3 folds the use of augmentations decreases the MSE on original designs in the hold-out sets on a model trained from scratch, with an average decrease of 16.5\%. In contrast, augmentations do not help pre-trained VAEs on original designs. We observe improvements using augmented designs for both VAE training cases in the reconstruction when including augmentated designs in the hold-out sets, pointing to the fact that training on augmentations makes the VAEs consistently more robust to perturbations (from a $0.033$/$0.041$ to $0.009$/$0.011$ MSE). We then compared training of a ViT-VAE from scratch or using pre-trained encoder weights as a warm-start. The reconstruction on hold-out sets is better in 3 out of 3 cases, training with ($15\pm 1\%$ better MSE) and without augmentations ($52\pm 7\%$ better). Augmentation partially substitutes pre-training, but it is still the second driver of hold-out set reconstruction improvement. The latent space constructed with pre-trained tranformer and augmentations serves as the basis for all subsequent optimization experiments.

\subsection{Transfer to nonlinear properties}

\subsubsection{Baselines}
As a gradient-based baseline, we use Convex Constrained Separable Approximations (CCSA), implemented by the NLOPT library ~\cite{NLopt} as the state of the art for constrained gradient-based inverse design and topology optimization. Note that here we are capable of deploying the gradient-based inverse designs thanks to previous work ~\cite{deng2022inverse} that constructed a fully-differentiable stress-strain curve adjoint simulator for this specific mechanical metamaterials platform (Supplementary Note~\ref{sec:platform-appendix}). In principle, the gradient might not be readily available, leading us to introduce also a gradient-free baseline. As a gradient free optimizer we deploy a CMA-ES, also implemented by NLOPT ~\cite{NLopt} and directly acting on the 324 dimensional space. After comparing their performance, we preferred CMA-ES to a Bayesian Optimizer (BO) as a gradient-free alternative on the full space as it performed slightly better. For the BO performed by TRAMA, we deploy a gaussian process surrogate with RBF kernel~\cite{williams2006gaussian} and an Upper Confidence Bound (UCB) acquisition function~\cite{srinivas2009gaussian}. In Supplementary Note~\ref{sec:opt-appendix}, Supplementary tables~\ref{tab:si-ccsa-tol} and~\ref{tab:si-cmaes-budget} justify the choice of the optimization hyperparameters of baselines.

\subsubsection{Transfer to stress-strain curve matching}
\label{sec:linear-curve}

To evaluate the transferability of our engineered search space, we first employ it for nonlinear stress-strain curve matching, going from a cheap linear task to computationally expensive nonlinear properties. Stress-strain curves encode the mechanical behavior of metamaterials across deformation regimes, making them a central design objective. Distinct features of the curve map directly onto functionality: the initial slope sets the stiffness, the area under the curve determines the energy absorbed, regions of negative slope enable snap-through instabilities and multistability. Targeting the full curve therefore allows designing for applications such as multistability, robotic actuation, and programmable or reconfigurable responses ~\cite{haghpanah2016multistable, florijn2014programmable, jin2020kirigami, chen2021reprogrammable, medina2020navigating}.

\noindent The Young's modulus, defined at an infinitesimal strain, evaluates the elastic response and therefore amounts to a linear system of equations. In contrast, a stress-strain curve depends on the entire response up to large deformation. At these strains, the geometry changes substantially as the structure deforms, through large rotations of the quadrilaterals and the onset of instabilities such as buckling, so the stiffness itself evolves with the deformation.

\noindent Using TRAMA and the baseline optimizers, we aim to find a design that matches a set of desired stress-strain curves. Formally, given $N$ equally spaced  points $\epsilon_1, ...\epsilon_N=\epsilon_{\max}$, for each curve we provide the desired values of the curve $f_1, ... f_N$. The objective is to find a metamaterials that matches the curve $f$, minimizing $\sum_{i=1}^N (f(\epsilon_i) - f_i)^2$. We perform experiments with three curves of growing difficulty: a \textit{linear} curve, a \textit{concave} curve, and, in Sec.~\ref{sec:snap-curve}, a \textit{snap} curve (one that grows linearly then drops sharply). We discuss the linear and concave curves here, since they are the primitives of a good energy absorption efficiency (Sec.~\ref{sec:energy}), and the snap curve at the end, as a harder case. To help visualization of error we report in the figure the following metric, similar to a curve fractional error:

\begin{equation}
    e(f) = \frac{1}{n}\sum_{i=1}^n \frac{(f(\epsilon_i) - f_i)^2}{f_i^2} \cdot 100
\end{equation}
 The full comparison of performances of linear and concave curves is in Table~\ref{tab:curve-matching}, we note that the difference in performance is within the margin of fabrication error of about $5\%$.
\noindent Linear stress-strain curves represent constant stiffness across the deformation range. Unsurprisingly, here TRAMA performs the best, having a great inductive bias coming from Young Modulus. TRAMA achieves a comparable error to CMA-ES ($\approx 1.00\%$ against the $3.12\%$ of CCSA), requiring approximately one order of magnitude fewer evaluations than CMA-ES and comparable cost to CCSA. (Tab.~\ref{tab:curve-matching}).

\noindent Concave stress-strain curves exhibit progressive compliance (decreasing stiffness, relaxing, with increasing strain). This is a property relatively common in the space of geometries and well represented in our training set (see Fig.~\ref{fig:garbage_combined}). These curves are one of the primitives that compose a good energy absorption efficiency, therefore are valuable for impact protection, and we expect good transfer since these geometries are already contained in the latent space. The optimizers achieve a similar error in the range of $1.63\%$ to $3.50\%$. TRAMA is approximately 1 order of magnitude faster than CMA-ES and also faster on average than CCSA once we account for forward and backward solves.

\begin{table}[t]
    \centering
    \small
    \setlength{\tabcolsep}{4pt}
    \begin{tabular}{lcccccc}
        \toprule
        & \multicolumn{3}{c}{Linear} & \multicolumn{3}{c}{Concave} \\
        \cmidrule(lr){2-4} \cmidrule(lr){5-7}
        Algorithm & Best & Mean $\pm$ std & Evaluations & Best & Mean $\pm$ std & Evaluations \\
        \midrule
        CCSA   & 3.12 & 3.68 $\pm$ 0.40 & $215 \pm 40$ & \textbf{1.63} & \textbf{2.94 $\pm$ 2.79} & 329 $\pm$ 200 \\
        CMA-ES & \textbf{0.98} & 2.18 $\pm$ 0.68 & $1500$ & 2.97 & 3.45 $\pm$ 0.55 & $1500$ \\
        TRAMA  & 1.00 & \textbf{1.93 $\pm$ 0.42} & \textbf{90 $\pm$ 28} & $3.00$ & $5.00$ $\pm$ 0.79 & \textbf{232 $\pm$ 51} \\
        \bottomrule
    \end{tabular}
    \caption{Transfer to linear and concave stress-strain curve matching. Errors are the metric $e(f)$ (\%); evaluations count both forward and backward passes as one. Mean $\pm$ std are over 10 runs for all optimizers. CMA-ES reached the fixed budget of 1500 evaluations without converging in every run. The best value in each column is in bold.}
    \label{tab:curve-matching}
\end{table}

\subsubsection{Transfer to Energy Absorption Efficiency Maximization}
\label{sec:energy}
We then employ the engineered search space for the maximization of energy absorption efficiency, which has the same computational expense as curve matching. Energy absorption efficiency is critical for applications that require impact mitigation and shock dissipation ~\cite{zhang2019energy, portela2021supersonic}. Given a stress-strain curve, the energy efficiency of a metamaterial measures the capacity to absorb energy normalized by the product of maximum stress and maximum strain experienced.
\[
\tau = \frac{\int_A f(\epsilon)  d\epsilon}{\max_{\epsilon \in A} \epsilon \max_{\epsilon \in A}f(\epsilon)}
\]

\begin{figure}
    \centering
    \includegraphics[width=0.8\linewidth]{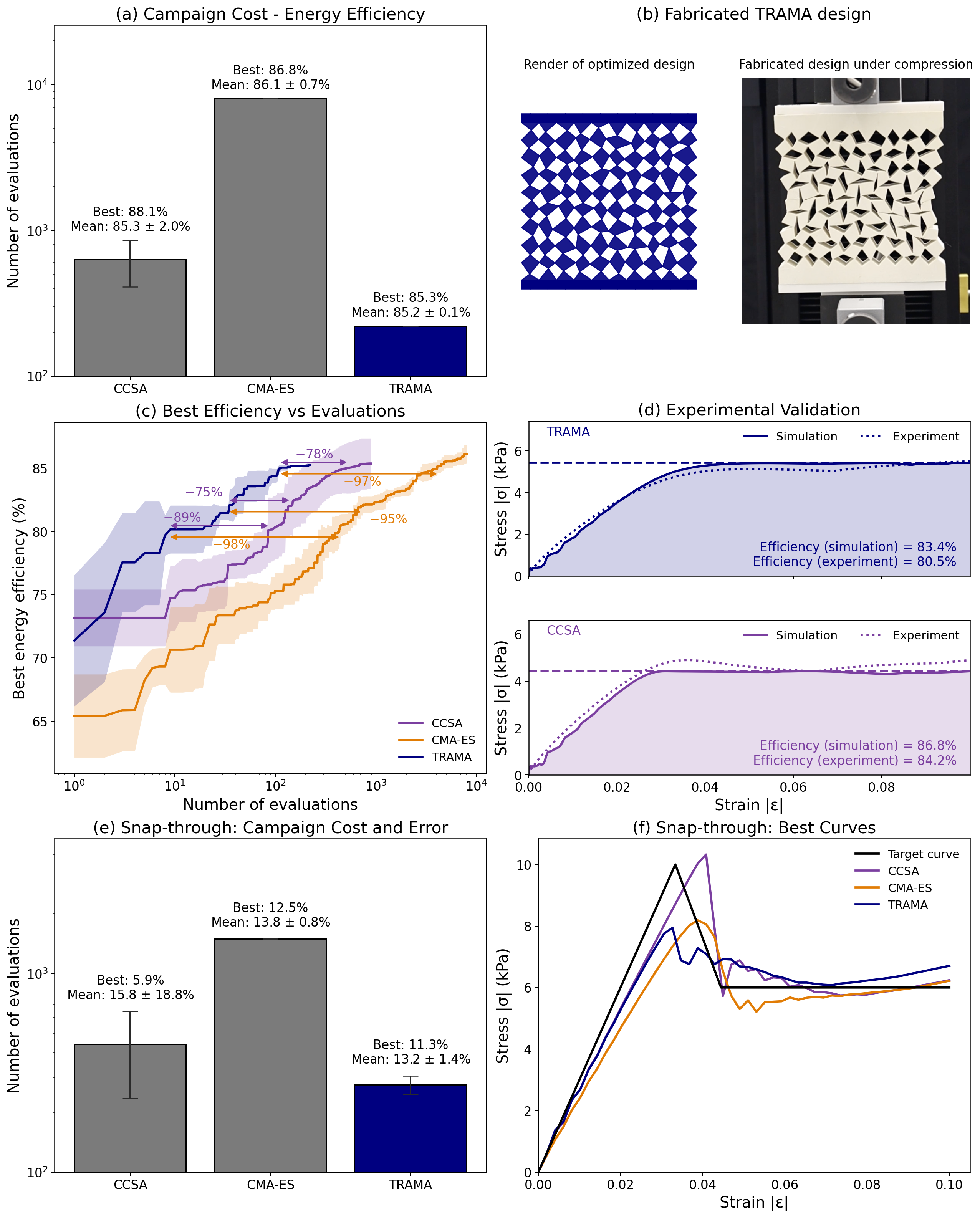}
    \caption{\textbf{Energy absorption efficiency maximization design and fabrication results} (a) TRAMA achieves a comparable average 85.2\% energy absorption efficiency with 35\% of CCSA's budget and with 3\% of CMA-ES's budget (b) TRAMA Optimized metamaterial geometry and fabricated design under compression. (c) Best design (mean $\pm$ standard deviation across 5 seeds) found by the optimizer for every number of evaluations. We show average budget savings to reach $80\%$, $82
    \%$ and $85\%$. (d) Simulated vs experimental stress-strain curves. The simulation-to-reality gap is $5.3\%$ overall ($5.9\%$ for TRAMA, $4.8\%$ for CCSA). (e-f) Young Modulus knowledge does not transfer as well to snap curve designs. CCSA is able to achieve best designs within 6\% of error. TRAMA consistently finds designs above $10$\% error, revealing latent space expressiveness limits.}
    \label{fig:energy}
\end{figure}

\noindent A stress-strain curve with high energy absorption efficiency combines a stiff initial linear phase with a long stress plateau. A long plateau holds the stress at a constant level, so that the absorbed energy, which is the area under the curve, accumulates at the maximum allowed stress over the entire available strain; the stiff initial linear phase brings the structure to this plateau stress almost immediately, so that little of the available strain is spent at low stress levels where little work is done. An ideal energy absorber is the limiting case of a perfectly flat plateau reached immediately. Intuitively, transfer will be successful because this mechanical response is related to physics that it is contained in the dataset and that we already found in the latent space, i.e. an initial responses encoded by Young's Modulus and a concave plateau region (see ~\ref{fig:garbage_combined}). Across 5 seeds, Figure~\ref{fig:energy}a shows the number of evaluations required by the baselines to achieve the energy-absorption efficiency of TRAMA, following the convention of counting both forward and backward passes per iteration. CMA-ES requires about 36 times more evaluations to get a design that performs comparable to TRAMA, with an higher per-seed variability. Gradient-based CCSA on average takes $631\pm 222$ evaluations to achieve what TRAMA does in $220$ forward evaluations (around $1/3$ of the computational expense). Fig.~\ref{fig:energy}c shows the average and standard deviation of the best efficiency found by each method with a given number of evaluations. TRAMA reaches every efficiency level up to $85$\% with $20–50$ times fewer evaluations than CMA-ES, corresponding to a consistent approximate $1.5$ order of magnitude speed up. Compared to CCSA, TRAMA achieves on average a given efficiency level with 3-5 times less evaluations. Being a local optimizer, CCSA is more dependent on the starting point of the optimization, resulting in a bigger final standard deviation. This means that, although performing similarly on average, its best designs are $1$ to $2$ points better than TRAMA's best designs. Fig.~\ref{fig:energy}b depicts an optimized design for TRAMA (not the best overall) and its fabricated version under compression. The interested reader can look into Supplementary Information for information about the fabrication (Supplementary Note~\ref{sec:fabrication}) and the experimental calibration (Supplementary Note~\ref{sec:exp-calibration}). In Fig.~\ref{fig:energy}d, we show the stress-strain curves of the fabricated and experimentally tested designs. The average simulation-to-reality gap, defined as the point-wise relative error $\|f_i^{\mathrm{sim}}-f^{\mathrm{exp}}_i\|_2/\|f_i^{\mathrm{exp}}\|_2$ between the simulated and measured stress-strain curves, is $5.3\%$ across all runs ($5.9\%$ for TRAMA and $4.8\%$ for CCSA). The measured efficiency is worse than the simulated one for both algorithms, with an average decrease of $3.1$ percentage points ($3.7$ for TRAMA and $2.5$ for CCSA). Repeating the schematic cost analysis we performed in Sec.~\ref{sec:methods} and using the tailored dynamical simulator, we obtain that computational savings are achieved on the first design for the gradient-free and after 6 designs against the gradient-based baseline. In such a case in which the design property is still relatively cheap, re-usability of the latent space for a new property can help ammortize the fixed-costs of dataset creation and training.

\subsubsection{Transfer to nonlinear stress-strain snap curve matching}
\label{sec:snap-curve}

We showcase the \textit{snap} curve as a harder case for which transfer fails (Fig.~\ref{fig:energy}e and Fig.~\ref{fig:energy}f). With a budget similar to the one used for other experiments, TRAMA underperforms with a best design $11.3\%$ error. In contrast, CCSA is able to effectively replicate the snap, achieving a best error of $5.9\%$. CMA-ES performs similarly to CCSA in terms of error, however, qualitatively the best CMA-ES design reproduces the drop in stress more faithfully than the best TRAMA design. As we demonstrate in the ablation analysis in Sec.~\ref{sec:ablation}, this performance gap stems from insufficient expressiveness in the engineered latent space. The space of optimal Young Modulus designs we used to create the latent space created a bias for the optimization of geometries that is misaligned with this snap behavior.

\subsection{Transferability Analysis of TRAMA}
\label{sec:ablation}

\begin{figure}[H]
    \centering
    \includegraphics[width=\linewidth]{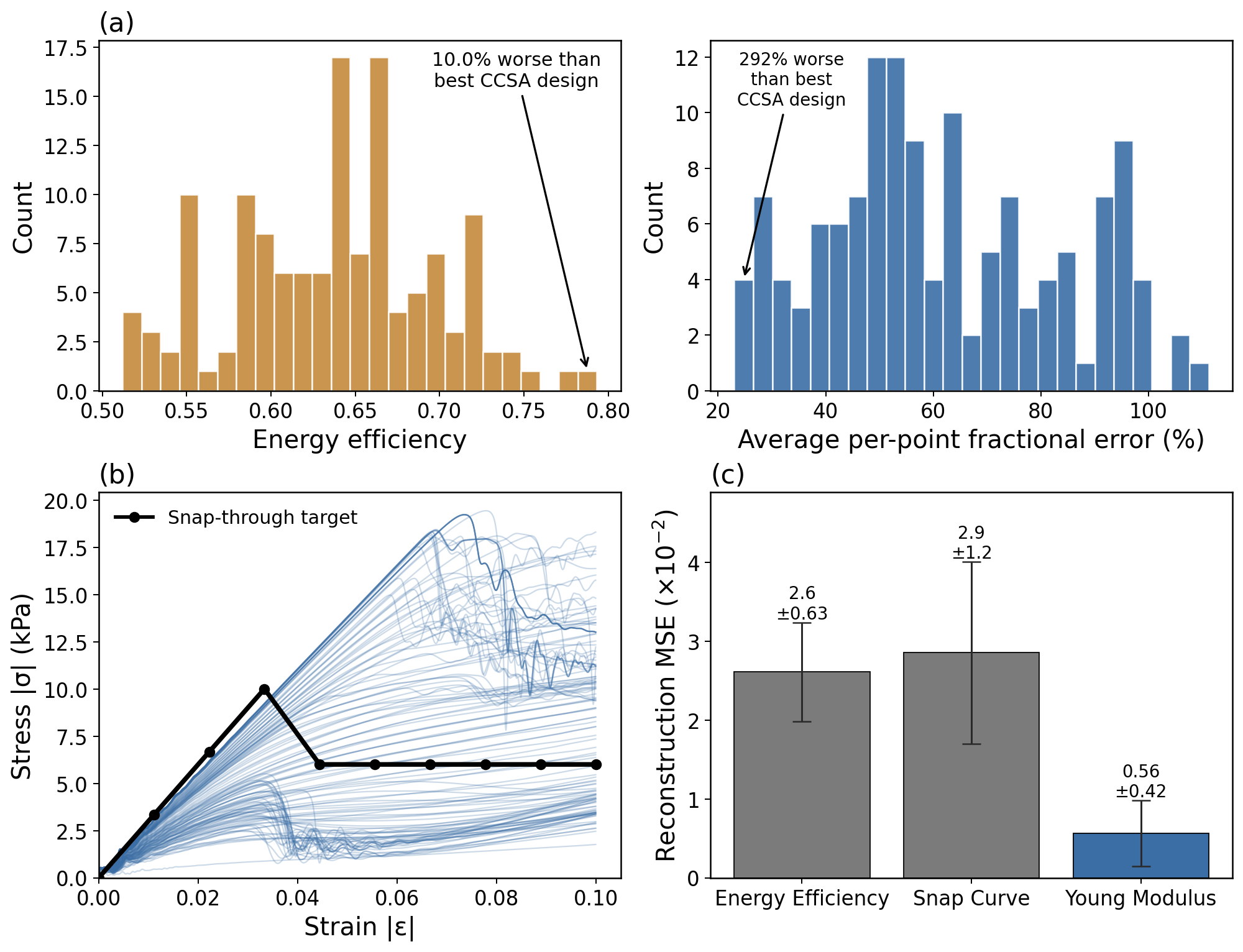}
    \caption{\textbf{TRAMA's training set transferability and quantitative out-of-distribution error} (a) The training set properties reveal misalignment between source and target design spaces. Left: Energy efficiency distribution shows competitive in-distribution designs at 79\%, only 9.9\% below CCSA's best. Right: Snap-curve fractional errors are 292\% worse than CCSA in the training set, indicating the target optimal structures lie outside the learned distribution. (b) Stress-strain curves for metamaterials in the training set. (c) Reconstruction error on CCSA's optimal designs for both properties is 4 to 5 times higher than validation error for Young's modulus. This is expected from training set misalignment for snap curve matching. It points to the fact gradient-based optimizers discover designs structurally distinct from those well-represented in the VAE's latent space for energy-absorption efficiency maximization.}
    \label{fig:garbage_combined}
\end{figure}

We analyze the transferability of TRAMA from a latent space constructed from Young's modulus to the two target properties: energy efficiency and snap curve matching. We showcase three possible ways of evaluating the transfer to a new property. Sec.~\ref{sec:energy} shows that the space successfully translated to the energy efficiency task, whereas it failed to recover designs that showcase the snap curve (Sec.~\ref{sec:snap-curve}). 
\\
\noindent The latent space is a learned representation of the geometric features that are present in the dataset. With our careful choice of data generation and augmentation, we nudge the representation to learn the features of high performing design for the original design task. Transfer is successful when the same underlying geometric features that guide Young's modulus optimization remain relevant for the new target property. Conversely, if the new problem depends on fundamentally different geometric primitives that are absent from the training data, the latent representation is unlikely to capture or express them effectively, even worse it may bias the design away from high-performing ones. In this sense, the first transferability test we perform is concerned with the alignment between the space of optimal Young's modulus designs and the space of optimal target property designs. Computing target properties for each design in the training set reveals that the energy efficiency distribution (Fig.~\ref{fig:garbage_combined}a) contains competitive designs at 79\% energy efficiency, only 9.9\% below CCSA's best result. By contrast, snap-curve designs require exploration outside the training distribution, with the best training-set error approximately 292\% worse than CCSA's best. Indeed, in Fig.~\ref{fig:garbage_combined}b, we showcase the stress-strain curves of designs in the training set. Curves that resemble the concave and linear case are present, while no snap curve with the same magnitude as the target appears. This test requires only the early-stopped optimization of the cheap property that builds the dataset, and the evaluation of the target property on a subset of its designs. Its cost is dominated by the training of the VAE, so it can be performed a priori, before committing to the construction of the latent space.\\

\begin{figure}[H]
    \centering
    \includegraphics[width=\linewidth]{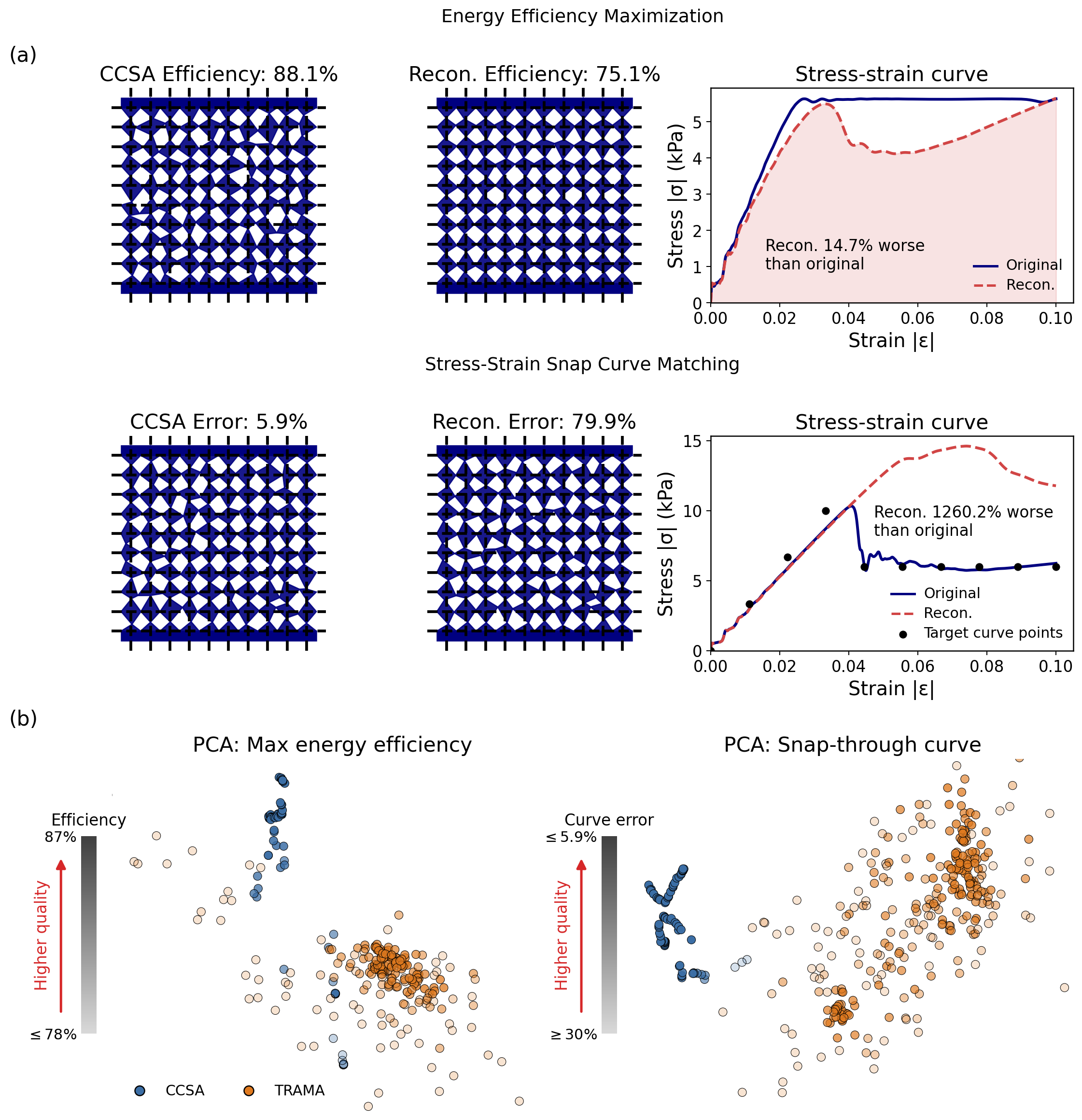}
    \caption{\textbf{TRAMA's qualitative out-of-distribution reconstruction and optimization trajectories} (a) Two qualitative examples of poor reconstruction of CCSA metamaterials by the VAE-ViT. On the left the designs found by CCSA for the two transferred properties. The reconstructed designs (center) have respectively $14.7\%$ and $1260.2\%$ worse metrics compared to the original structures. (b) PCA projections of multiple CCSA campaigns (blue) and the best TRAMA campaign (orange). The lower the quality in terms of the metric of interest, the higher the transparency. The total variance captured is respectively $32.5\%$ and $34.9\%$.}
    \label{fig:recon_pca}
\end{figure}

\noindent The second test examines the decoder's capacity to reconstruct optimal designs for the transferred property after training. Specifically, we check whether the trained autoencoder reconstructs the optimal designs found by CCSA. If optimal designs are absent from the training set, then CCSA solutions are likely not representable in the learned latent space. Reasonably, this is the case of the snap curve experiment. As we observe from Fig.~\ref{fig:garbage_combined}c, the mean reconstruction error for $5$ CCSA optimal designs is around $4$ to $5$ times bigger than the one for optimal Young Modulus designs (taken from the test set of the ViT-VAE). Surprisingly, we notice a similar result in the case of the $5$ CCSA designs for energy efficiency. This suggests that good designs in the training set differ structurally from those found by CCSA, highlighting the value of generating the dataset with multiple optimizers in our dataset generation strategy. In Fig.~\ref{fig:recon_pca}a, we showcase qualitatively the effect of poor reconstruction for the best designs found by CCSA for the two tasks (left-most column). The central column showcases the two designs after reconstruction by the autoencoder. On these two examples, the bad reconstruction yields objectives that are $14.7$\% and $1260.2$\% worse than the original designs, respectively for the energy efficiency and snap curve matching.\\
\\

\noindent Finally we analyze how the latent space from the decoder is explored during the optimization and how it compares with CCSA runs. Even when the latent space contains in-distribution solutions and reconstruction performs adequately, the Bayesian optimizer may struggle to locate them efficiently. To investigate this, we project the geometries explored by TRAMA and CCSA onto the first two principal components of the design space (Fig.~\ref{fig:recon_pca}b), encoding solution quality via point transparency. Details about the fitting of the PCA are reported in Supplementary Note~\ref{sec:exp-appendix}. The quality of the solutions is encoded by the transparency of the points (the better the quality, the less transparent). The space is high dimensional and the first principal components capture only around 30\% of the diversity of optimal designs; nonetheless, the PCA analysis confirms that TRAMA totally skips the regions with good CCSA designs since they are out-of-distribution with respect to its training set. The second and third tests require a trained VAE and an optimization in the low-dimensional latent space, and are therefore a posteriori: they explain why a transfer underperforms rather than anticipate it.

\section{Discussion}

\noindent We present empirical evidence of linear-to-nonlinear transferability using TRAMA. This work adds to the broader literature that aims to enable previously unreachable expensive design solutions, and paves the way for research directions that remain open. TRAMA may extend to other properties in mechanics and other fields of PDE-constrained optimization. The components of the methods tailored particularly to our architected materials should apply directly to other choices of parametrizations. Many design tasks in science and engineering pair cheap and mathematically well-behaved physics with expensive and potentially non-differentiable ones. The cheap physics informs the geometrical features of the good designs and the resulting low-dimensional representation enables any optimizer, differentiable or not, to explore the expensive problem. In this view, the learned latent space is a reusable asset that is built once and amortized as it transfers across multiple downstream objectives. In contrast, physical-response surrogates are tied to a single target. The same logic applies wherever a low-cost proxy exists, for instance from static to dynamic responses~\cite{gongora2022designing}, from single- to multi-physics~\cite{chakraborty2021transfer}, or from low to high-fidelity simulators~\cite{pestourie2023physics, varagnolo2026physics}. As showcase by the snap example, a static latent space also has limitations. In the future, the finetuning of the latent space could better capture the diversity of optimal geometries for a new, poorly-transferrable, property. The transferability could be diagnosed a priori to determine whether further data generation is needed for the new property to avoid failed transfer attempts. Our post-hoc transferability analysis takes a step towards that by proposing a multi-step diagnostic framework to better understand our method and its transferability. Our choice to start with Young modulus was driven by convenience, and the transfer to energy maximization was motivated by field expertise and confirmed by our empirical results. However, future work could inform the choice of curriculum of transferable properties more systematically and in a physics-agnostic manner.

\noindent Another direction of research concerns the interaction between the latent space structure and the optimization dynamics. The KL-divergence weight during training controls latent space smoothness and structure, yet its relationship to optimizer behavior and performance should be further investigated. Extension applications with fully 3D parametrized materials promises even larger computational gains from the low-dimensional latent space because the parameterized dimensionality grows exponentially. The simulation costs will also increase the gap between the cheap proxy and the target nonlinear property.

\section{Methods}
\label{sec:methods}

The TRAMA (TRAnsferable representation learning for Metamaterial Architectures) framework  is applied here to the linear-to-nonlinear transfer of design knowledge for parametrized mechanical metamaterials. The hybrid idea is to combine a library of gradient‑based optimizations with a generative variational autoencoder (VAE) and a Bayesian optimizer. The framework operates in three stages: First, a library of designs with diverse geometric features is created using gradient‑based optimizations on inexpensive linear properties such as the Young modulus. Although not optimal, these designs contain transferable geometric primitives and therefore encode useful inductive bias in the optimization of other more complex properties of energy maximization and strain-stress curve matching. These designs are further augmented with no additional simulations using field-domain heurisitics. Secondly, a VAE compresses these designs into a compact latent space, reducing dimensionality by orders of magnitude, enabling gradient-free optimizer in this new low-dimensional latent space rather than on the high-dimensional parametrization. Third, an off‑the‑shelf Bayesian optimizer efficiently searches this low‑dimensional latent space for structures optimizing more computationally expensive or non‑differentiable properties. 

\begin{figure}[H]
    \centering
    \includegraphics[width=1\linewidth]{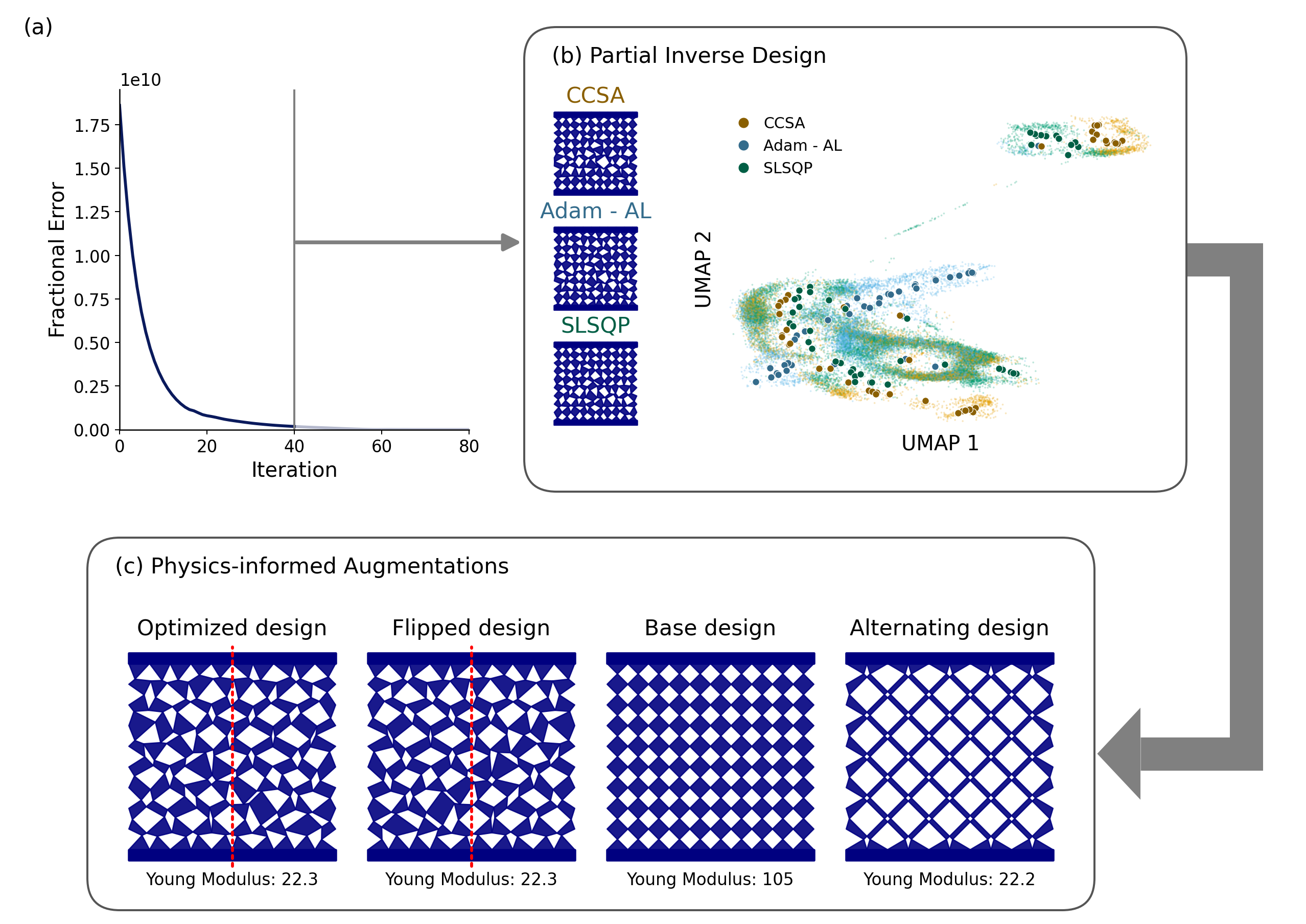}
    \caption{\textbf{TRAMA's dataset generation pipeline} (a) The pipeline for dataset creation, comprising of early stopped inverse design using a diverse set of optimizers, and physically-informed augmentations. (b) UMAP of the generated dataset, colored by optimizer used. Darker dots are associated with original datasets, while lighter smaller dots with the variants of those designs. ADAM-AL are distinguishable from designs from CCSA and SLSQP designs, while it is harder to separate the latter two. This is reasonable since they use similar surrogate models. (c) Example of augmentations that are invariant or diminishing the design Young Modulus. This yield a $\times 235$ increase in size of the dataset. }
    \label{fig:augm}
\end{figure}

\paragraph{Generation of a diverse initial dataset with early stops on inverse design runs}
Firstly, we choose an inexpensive and differentiable property to create the dataset. Early-stopped inverse‑design runs are performed using different gradient‑based optimizers, similar to~\cite{marzban2025hilab}. Our approach differs in two significant ways. First, the use of different optimizers promotes diversity in the resulting designs (Fig.~\ref{fig:augm}). Second, the number of iterations was chosen looking at the average error with respect to the iteration number, stopping before diminishing returns in the optimization curves (as depicted in Fig~\ref{fig:augm}).

\paragraph{Data augmentation with physically informed transformations}
To enrich this library without additional simulations, each design is modified by applying a set of physical rules that alter significantly the geometry while maintaining a good performance on the task they were optimized for. The transformations, and compositions of multiple of them, allow us to augment the dataset of around $230$ times, preserving physical meaningfulness and instilling useful domain-expert knowledge in the dataset. This includes mirror symmetries (horizontal, vertical, and 180 degree rotations), translations or parametrization-specific transformations, as shown in Fig.~\ref{fig:augm}. If a transformation drives the geometry outside the fabricable space (i.e. the set of designs satisfying parametrization-specific fabrication requirements) then the design is filtered out of the dataset.

\paragraph{Pre-trained Image Model}
We construct the latent space with a VAE. Input and reconstructed outputs are determined by the representation choice for the design, mainly images of the design itself or parameters that define the geometry as in the case of architected materials. The choice of the architecture depends on the high-dimensional input we provide, where convolutional neural networks (CNNs) will be preferred for images, while multi-layer perceptrons (MLPs) would work for parametrizations vectors. When the input is an image, the encoder benefits from pre-trained models on large generic image datasets. Such pre-trained representations have been shown to reduce the amount of task-specific data required~\cite{yosinski2014transferable, kornblith2019better}, which is particularly valuable here as we focus on the computational efficiency of our pipeline. Several encoder architectures are possible, and we adopt Vision Transformers (ViT)~\cite{dosovitskiy2021image}, similarly to Ref.~\cite{marzban2025hilab}. Here, the encoder takes a rendering of the architected lattice as a single image and pools the pre-trained ViT patch tokens with a learned attention to obtain the latent space. The decoder, in contrast to a standard image autoencoder, does not reconstruct the image: two MLPs output the nodal displacements on the horizontal and vertical axes of the lattice, so that decoded samples are directly geometries whose feasibility can be evaluated exactly. Training minimizes the reconstruction error, which forces the latent vector to retain the information needed to rebuild the design, plus a Kullback-Leibler (KL) divergence to a standard normal prior, which encourages a smooth, well-structured latent space in which nearby points decode to similar designs. Further architectural details are given in the Supplementary Note~\ref{sec:exp-appendix}.

\paragraph{Optimization in the latent-physical space}
Once the VAE is trained, its decoder $\mathcal{D}$ is frozen and used as a design generator that maps the low-dimensional latent space to the parameter space $\mathcal{D}(\mathbf{z})=\mathbf{p}$. We thereby reformulate the constrained design problem in original parameter space as a low-dimensional unconstrained problem in the latent space,
\begin{equation}
\begin{aligned}
\min_{\mathbf{p}\in\mathbb{R}^{n_p}} \;& J(\mathbf{u};\mathbf{p}) \\
\text{s.t.}\;& \mathbf{F}(\mathbf{p})\le\mathbf{0}
\end{aligned}
\qquad\longrightarrow\qquad
\min_{\mathbf{z}\in\mathcal{Z}} \; \tilde{J}(\mathbf{z}) = J\big(\mathbf{u};\mathcal{D}(\mathbf{z})\big)
+ \lambda\, G\big(\mathcal{D}(\mathbf{z})\big),
\label{eq:latent-opt}
\end{equation}
where $\mathbf{p}$ is the vector of geometric parameters, $J$ is the expensive, potentially non-differentiable
objective (e.g., energy absorption efficiency or stress-strain curve mismatch), $\mathbf{u}$ is the mechanical
response of design $\mathbf{p}$ obtained from the simulator, and $\mathbf{F}(\mathbf{p})\le\mathbf{0}$ collects the
fabricability constraints, which depend on the geometric parameters only. In the latent space the constraints
enter the objective as a penalty: $G(\mathbf{p}) = \max_i F_i(\mathbf{p})$ if this violation exceeds a
tolerance and $G=0$ otherwise, weighted by the coefficient $\lambda \ge 0$. Here the optimizer is chosen to be a Bayesian Optimizer (BO) as in Ref.~\cite{marzban2025hilab} but any off-the-shelf method can be used, including gradient-based ones if the function we optimize has a differentiable solver available. 

\paragraph{Computational Benefits}

We provide an example outline of the computational savings enabled by our method. The cost of our method splits into a fixed, one-time part (dataset construction and VAE training) and a per-campaign part (Bayesian optimization in the latent space). We showcase the computational benefit of this approach using the cost of 2D FEM simulations on ABAQUS, the alternative if a tailored dynamic solver is not available. Let $c_N \approx 14$~min be the cost of one evaluation of the nonlinear property (e.g. the stress-strain curve), and $c_L \approx 1.6$~ms that of the cheap linear property (e.g. Young Modulus, for information on our linearized implementation see Supplementary Note~\ref{sec:linearized-modulus}), so that $c_N/c_L \approx 525000$. We optimize $N_0 = 130$ designs for the Young's modulus, stopping at $K = 40$ iterations instead of the $\sim 90$ needed for full convergence. The cost of the dataset is
\begin{equation}
C_{\mathrm{data}} = N_0\, K\, c_L \approx 130 \cdot 40 \cdot 1.6\times10^{-3}~\mathrm{s} \approx 8~\mathrm{s}.
\end{equation}
Physically meaningful augmentation expands the dataset to $N_{\mathrm{aug}} = 3\times10^4$ designs without any additional simulation. Producing the same designs by optimization would instead cost $N_{\mathrm{aug}} K c_L \approx 32$~min with the cheap property. With the GPU-accelerated VAE training time $T_{\mathrm{train}} \approx 20$~min, the total fixed cost of our strategy amounts to
\begin{equation}
C_{\mathrm{fixed}} = C_{\mathrm{data}} + T_{\mathrm{train}} \approx 1.2\times10^{3}~\mathrm{s} \approx 20~\mathrm{min},
\end{equation}
Bayesian optimization in the latent space uses $n_{\mathrm{BO}} \approx 220$ evaluations of the expensive property, and the cost of fitting the Gaussian process is negligible in comparison:
\begin{equation}
C_{\mathrm{BO}} = n_{\mathrm{BO}}\, c_N \approx 50~\mathrm{h}.
\end{equation}
The alternatives are fully converged gradient-based optimization in the original parameter space, with $n_{\mathrm{GB}} \approx 630$ evaluations for energy absorption efficiency, or gradient-free optimization in the original space (CMA-ES), with $n_{\mathrm{GF}} \approx 8000$ evaluations:
\begin{equation}
C_{\mathrm{GB}} = n_{\mathrm{GB}}\, c_N \approx 147~\mathrm{h}, \qquad C_{\mathrm{GF}} = n_{\mathrm{GF}}\, c_N \approx 1867~\mathrm{h}.
\end{equation}
The latent space is built once and reused for every new property or target, so $M$ campaigns cost $C_{\mathrm{fixed}} + M\,C_{\mathrm{BO}}$ with our method, against $M\,C_{\mathrm{GB}}$ or $M\,C_{\mathrm{GF}}$ for the baselines. Under this realistic conditions, our method is cheaper from the very first design, and can be reused for other properties. Potentially, the BO loop could be run twice as long and our method would still be cheaper. In the rest of the study, we use the differentiable dynamics simulator developed in Ref.~\cite{deng2022inverse} and described in the Supplementary Note~\ref{sec:platform-appendix}, without loss of generality for the empirical results on transferability. Note that the error that separates the 2D FEM and the deployed dynamical simulator is smaller than the simulation-reality gap of the 2D FEM.

\section*{Data availability}
The datasets and trained model weights generated in this study have been deposited in Zenodo~\cite{varagnolo_2026_zenodo}. They are available to reviewers during peer review and will be made publicly available upon publication..

\section*{Code availability}
The code and reproducibility instructions to generate our results is provided to reviewers and will be released in a Github repository~\cite{varagnolo2026trama_code} upon publication.

\bibliography{sn-bibliography}

\section*{Funding}
The authors discloses support for the research of this work from the National Science Foundation [grant number IIS-2435905].

\section*{Author Contributions}
A.V. and M.D. contributed equally to this work. M.D. developed the code the linearized young modulus solver. A.V. developed the code for the machine learning model, the inverse design experiments and the transferability analysis.  M.D. carried out the experiments on fabricated designs. B.D. and R.P. conceived and supervised the project and acquired funding. A.V. and M.D. wrote the original draft; all authors reviewed and edited the manuscript.

\section*{Competing Interests}
The authors declare no competing interests.

\ifwithSI
\clearpage
\makeatletter
\setcounter{section}{0}\setcounter{figure}{0}\setcounter{table}{0}
\setcounter{secnumdepth}{1}
\renewcommand{\thesection}{\arabic{section}}
\def\@seccntformat#1{Supplementary Note \csname the#1\endcsname:\hskip.5em}
\renewcommand{\figurename}{Supplementary Fig.}
\renewcommand{\tablename}{Supplementary Table}
\numberwithin{equation}{section}
\def\theHsection{SI.\arabic{section}}
\def\theHfigure{SI.\arabic{figure}}
\def\theHtable{SI.\arabic{table}}
\def\theHequation{SI.\arabic{section}.\arabic{equation}}
\makeatother
\section*{Supplementary Information}
\section{Mechanical Metamaterials Simulation and Experimental Platform}
\label{sec:platform-appendix}
\subsection{Design Space}
\label{sec:design-space}

\quad
The metamaterial consists of an $N_x\times N_y$ array of quadrilaterals. Here, $a$ denotes the diagonal length of an unperturbed square (Supplementary Figure~\ref{fig:design-space}(a)). Rows and columns are indexed from bottom to top and from left to right, respectively. For each quadrilateral $[i,j]$, the local index $p=1,2,3,4$ denotes its right, upper, left, and lower vertices, respectively.

Each vertex is displaced from its original position by a perturbation vector $\mathbf d_p^{[i,j]}=[d_{p,x}^{[i,j]},d_{p,y}^{[i,j]}]^{\mathsf T}$. The position vectors from the center of the unperturbed square to its four perturbed vertices are given by
\begin{equation}
\begin{aligned}
\mathbf r_1^{[i,j]}&=
\begin{bmatrix}
a/2\\
0
\end{bmatrix}
+\mathbf d_1^{[i,j]},
&
\mathbf r_2^{[i,j]}&=
\begin{bmatrix}
0\\
a/2
\end{bmatrix}
+\mathbf d_2^{[i,j]},\\
\mathbf r_3^{[i,j]}&=
\begin{bmatrix}
-a/2\\
0
\end{bmatrix}
+\mathbf d_3^{[i,j]},
&
\mathbf r_4^{[i,j]}&=
\begin{bmatrix}
0\\
-a/2
\end{bmatrix}
+\mathbf d_4^{[i,j]}.
\end{aligned}
\label{eq:dm-reference-geometry}
\end{equation}

Because each vertex is shared by two neighboring quadrilaterals, the corresponding perturbation vectors satisfy
\begin{equation}
\mathbf d_1^{[i,j]}=\mathbf d_3^{[i,j+1]},\qquad \mathbf d_2^{[i,j]}=\mathbf d_4^{[i+1,j]}.
\label{eq:dm-shared-offsets}
\end{equation}

Each independent perturbation vector is constrained to the diamond-shaped design region
\begin{equation}
\left\|\mathbf d_p^{[i,j]}\right\|_1\leq\frac{a}{2}.
\label{eq:dm-design-boundary}
\end{equation}
Perturbations outside this region are rejected and resampled (Supplementary Figure~\ref{fig:design-space} (b)).
\begin{figure}[htbp]
    \centering
    \includegraphics[width=0.8\linewidth]{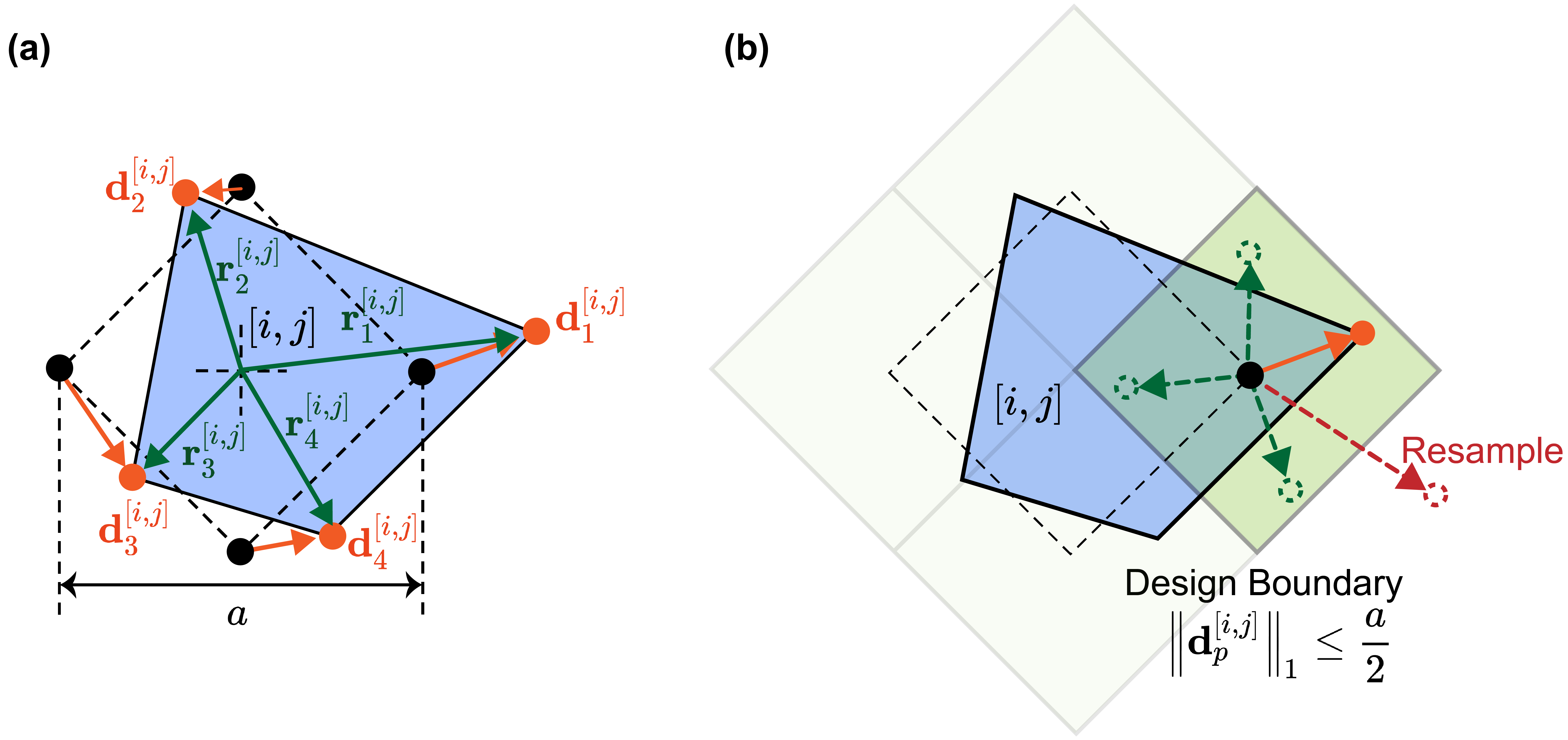}
    \caption{Geometric design space. (a) Perturbation vectors $\mathbf d_p^{[i,j]}$ and position vectors $\mathbf r_p^{[i,j]}$ of the quadrilateral vertices. (b) Diamond-shaped design region for each perturbation vector.}
    \label{fig:design-space}
\end{figure}

\newpage
\subsection{Fabrication and Experiments}

\subsubsection{Fabrication}
\label{sec:fabrication}
\quad
The physical samples are based on a $10\times10$ array of unperturbed squares with a diagonal length of $a=10~\mathrm{mm}$. The unperturbed structure has planar dimensions of $100~\mathrm{mm}\times100~\mathrm{mm}$ and an out-of-plane thickness of $20~\mathrm{mm}$. The connecting ligaments have an in-plane thickness of $1~\mathrm{mm}$. The disordered designs are generated by displacing the vertices according to the sampled perturbation vectors.

The samples are 3D printed by Inkbit using TEPU\texttrademark~30A, a soft elastomer with a Shore A hardness of 32 and an elongation at break of 200\%. This material is selected to accommodate the large local deformations of the ligaments during compression.

\subsubsection{Compression Tests}

\quad
Compression tests were performed using an Instron universal testing machine (Model 68SC-1, Instron, USA). The samples were compressed under displacement control at a constant crosshead speed of $3~\mathrm{mm/s}$ to a compressive strain of $-10\%$. This corresponds to a displacement of $9~\mathrm{mm}$, based on the initial $90~\mathrm{mm}$ distance between the centers of the top- and bottom-row quadrilaterals. Each sample underwent three consecutive compression cycles. The first cycle was used for mechanical preconditioning and excluded from the analysis. The force-displacement responses from the subsequent two cycles were averaged for analysis.

\newpage
\subsection{Discrete Model}
\label{sec:app-model-discrete}

\quad
\begin{figure}[htbp]
    \centering
    \includegraphics[width=1\linewidth]{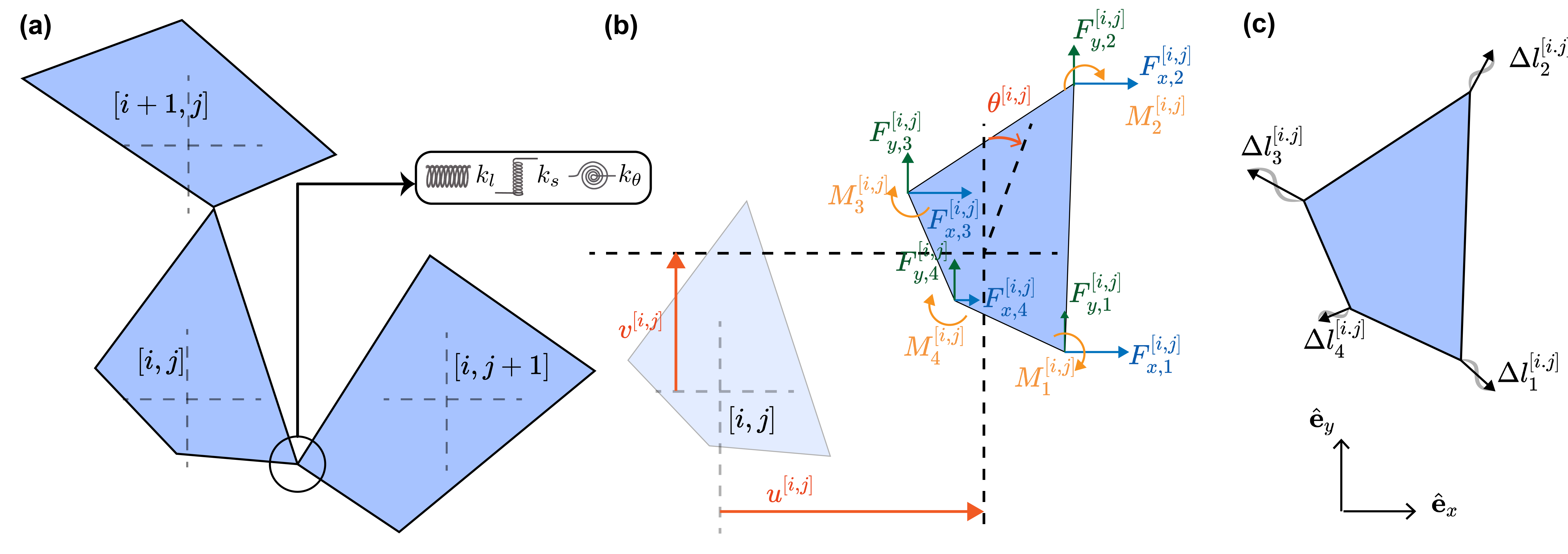}
    \caption{Discrete mechanical model. (a) Rigid quadrilaterals connected by ligaments, each represented by longitudinal, shear, and torsional springs. (b) In-plane translations and rotation of a quadrilateral, together with the forces and moments acting on it. (c) Spring displacement vectors $\Delta\mathbf l_p^{[i,j]}$ for the four ligaments connected to quadrilateral $[i,j]$.}
    \label{fig:discrete-model}
\end{figure}

The quadrilaterals are treated as rigid bodies connected at their vertices by ligaments. Each quadrilateral $[i,j]$ has three degrees of freedom, collected in the DOF vector
\begin{equation}
\mathbf q^{[i,j]}=
\begin{bmatrix}
u^{[i,j]}\\
v^{[i,j]}\\
\theta^{[i,j]}
\end{bmatrix},
\label{eq:dm-dof}
\end{equation}
where $u^{[i,j]}$ and $v^{[i,j]}$ are the translations along the $x$- and $y$-directions, and $\theta^{[i,j]}$ is the in-plane rotation angle, with clockwise rotations taken as positive (Supplementary Figure~\ref{fig:discrete-model} (b)). Using the vertex numbering defined in the Design Space section, the relative rotation at each connection is defined as the rotation of quadrilateral $[i,j]$ minus that of its neighboring quadrilateral:
\begin{equation}
    \begin{bmatrix}
        \Delta\theta_1^{[i,j]}\\
        \Delta\theta_2^{[i,j]}\\
        \Delta\theta_3^{[i,j]}\\
        \Delta\theta_4^{[i,j]}
    \end{bmatrix}
    =
    \begin{bmatrix}
        \theta^{[i,j]}-\theta^{[i,j+1]}\\
        \theta^{[i,j]}-\theta^{[i+1,j]}\\
        \theta^{[i,j]}-\theta^{[i,j-1]}\\
        \theta^{[i,j]}-\theta^{[i-1,j]}
    \end{bmatrix}.
    \label{eq:dm-relative-rotation}
\end{equation} 

As shown in Supplementary Figure~\ref{fig:discrete-model} (a), each ligament is represented by a longitudinal spring with stiffness $k_l$, a shear spring with stiffness $k_s$, and a nonlinear torsional spring, which exerts the moment
\begin{equation}
    M_{\theta,p}^{[i,j]}
    =
    -k_\theta
    \left[
        \Delta\theta_p^{[i,j]}
        +\gamma\left(\Delta\theta_p^{[i,j]}\right)^3
    \right],
    \label{eq:dm-torsion}
\end{equation}
where $k_\theta$ is the torsional stiffness and $\gamma$ is a dimensionless parameter controlling the nonlinear rotational response. For simplicity, the longitudinal and shear springs are assumed to remain parallel to either the global $x$- or $y$-axis. All ligaments share the same stiffness parameters.

Contact between neighboring quadrilaterals is modeled using an additional rotational spring. At each connection $p$ of quadrilateral $[i,j]$, two pairs of edges may come into contact. Their angular gaps are denoted by $\beta_{\mathrm{I},p}^{[i,j]}$ and $\beta_{\mathrm{II},p}^{[i,j]}$ (Supplementary Figure~\ref{fig:contact}). As the neighboring quadrilaterals rotate relative to each other, one gap increases while the other decreases. Contact is activated when either angular gap falls below the threshold $\beta_c$.

The contact stiffness is defined as
\begin{equation}
K_{\mathrm{cont}}(\beta)=
\begin{cases}
k_{\mathrm{cont}}, & \beta<\beta_c,\\
0, & \beta\geq\beta_c,
\end{cases}
\label{eq:dm-contact-stiffness}
\end{equation}
where $k_{\mathrm{cont}}$ is the contact stiffness in the active state and $\beta_c$ is the activation threshold. The contact moment acting on quadrilateral $[i,j]$ at vertex $p$ is
\begin{equation}
M_{\mathrm{cont},p}^{[i,j]}=
\left(\beta_{\mathrm{I},p}^{[i,j]}-\beta_c\right)K_{\mathrm{cont}}\!\left(\beta_{\mathrm{I},p}^{[i,j]}\right)
-\left(\beta_{\mathrm{II},p}^{[i,j]}-\beta_c\right)K_{\mathrm{cont}}\!\left(\beta_{\mathrm{II},p}^{[i,j]}\right).
\label{eq:dm-contact-moment}
\end{equation}

Following the vertex numbering defined in the Design Space section, a horizontal ligament connects vertex $p=1$ of quadrilateral $[i,j]$ to vertex $p=3$ of quadrilateral $[i,j+1]$, while a vertical ligament connects vertex $p=2$ of quadrilateral $[i,j]$ to vertex $p=4$ of quadrilateral $[i+1,j]$ (Supplementary Figure~\ref{fig:contact}). The contact moments acting on the two neighboring quadrilaterals are equal and opposite:
\begin{equation}
\begin{aligned}
    M_{\mathrm{cont},1}^{[i,j]}&=-M_{\mathrm{cont},3}^{[i,j+1]},\\
    M_{\mathrm{cont},2}^{[i,j]}&=-M_{\mathrm{cont},4}^{[i+1,j]}.
\end{aligned}
\label{eq:dm-contact-action-reaction}
\end{equation}

\begin{figure}
    \centering
    \includegraphics[width=0.9\linewidth]{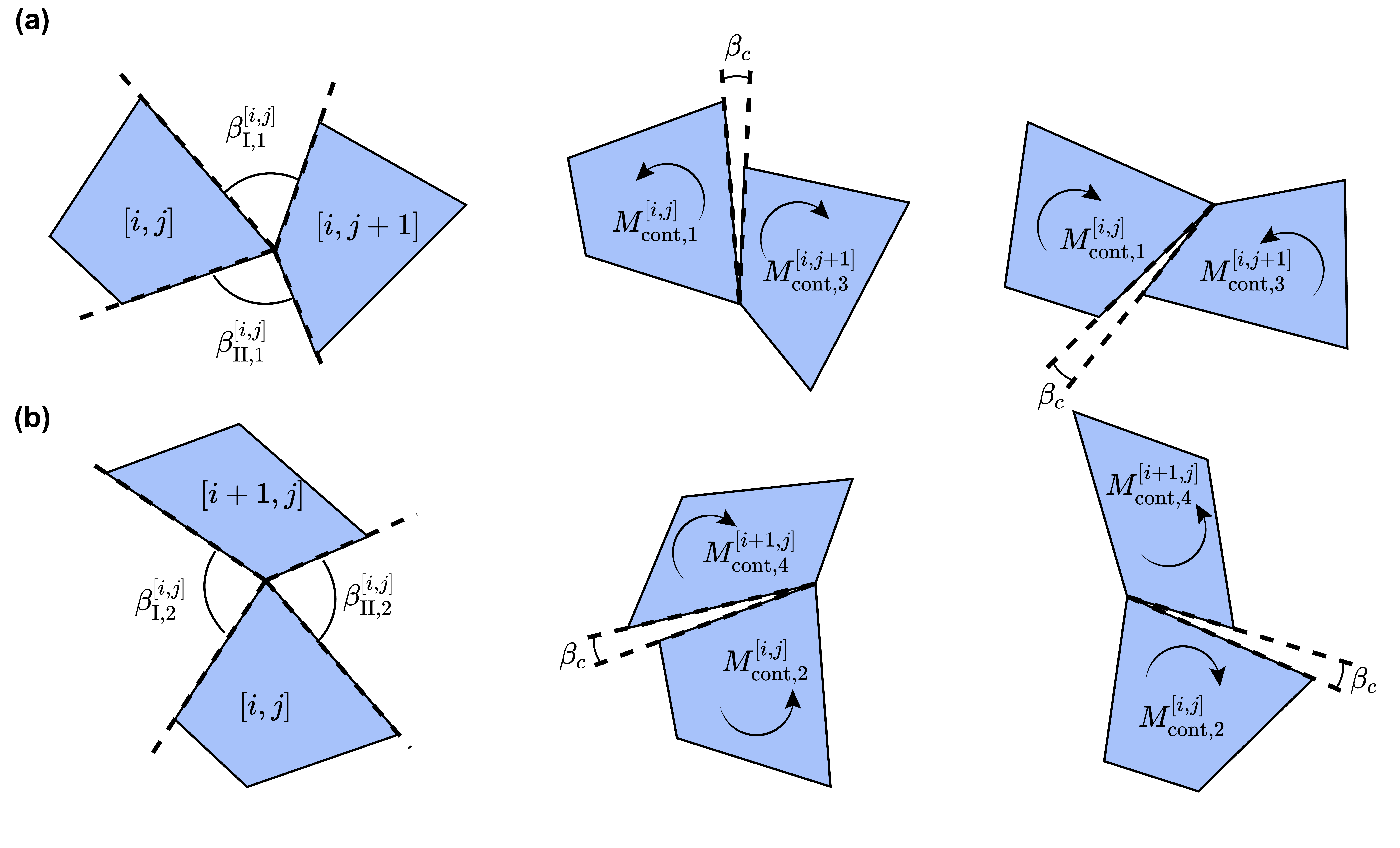}
    \caption{Contact model for (a) horizontally and (b) vertically adjacent quadrilaterals. When either angular gap falls below $\beta_c$, the contact spring generates equal and opposite moments on the neighboring quadrilaterals.}
    \label{fig:contact}
\end{figure}

The equations of motion for each quadrilateral(Supplementary Figure~\ref{fig:discrete-model}(b)) are
\begin{equation}
\begin{aligned}
m\ddot u^{[i,j]}&=\sum_{p=1}^{4}F_{x,p}^{[i,j]},\\
m\ddot v^{[i,j]}&=\sum_{p=1}^{4}F_{y,p}^{[i,j]},\\
J\ddot\theta^{[i,j]}&=\sum_{p=1}^{4}M_p^{[i,j]}.
\end{aligned}
\label{eq:dm-dynamics}
\end{equation}
where $m$ and $J$ are the mass and moment of inertia of each quadrilateral, respectively. The force components and total moment acting on quadrilateral $[i,j]$ associated with its $p$-th vertex are
\begin{equation}
\begin{aligned}
    F_{x,p}^{[i,j]}
    &=
    \left(
        \mathbf k_p\Delta\mathbf l_p^{[i,j]}
    \right)\cdot\hat{\mathbf e}_x,\\
    F_{y,p}^{[i,j]}
    &=
    \left(
        \mathbf k_p\Delta\mathbf l_p^{[i,j]}
    \right)\cdot\hat{\mathbf e}_y,\\
    M_p^{[i,j]}
    &=
    -
    \left[
        \mathbf R\left(\theta^{[i,j]}\right)
        \mathbf r_p^{[i,j]}
        \times
        \left(
            \mathbf k_p\Delta\mathbf l_p^{[i,j]}
        \right)
    \right]\cdot\hat{\mathbf e}_z
    +M_{\theta,p}^{[i,j]}
    +M_{\mathrm{cont},p}^{[i,j]}.
\end{aligned}
\label{eq:dm-forces-moments}
\end{equation}

Here, $\hat{\mathbf e}_x$, $\hat{\mathbf e}_y$ and $\hat{\mathbf e}_z$ are the Cartesian unit vectors, with $\hat{\mathbf e}_z$ pointing out of the plane. The stiffness matrix $\mathbf k_p$ is defined as
\begin{equation}
    \mathbf k_p=
    \begin{cases}
        \begin{bmatrix}
            k_l&0\\
            0&k_s
        \end{bmatrix},
        &p=1,3,\\[8pt]
        \begin{bmatrix}
            k_s&0\\
            0&k_l
        \end{bmatrix},
        &p=2,4.
    \end{cases}
    \label{eq:dm-stiffness-matrix}
\end{equation}

The rotation matrix $\mathbf R(\theta)$ describing the in-plane rotation of each quadrilateral is defined as
\begin{equation}
    \mathbf R(\theta)=
    \begin{bmatrix} 
        \cos\theta & \sin\theta\\
        -\sin\theta & \cos\theta
    \end{bmatrix}.
    \label{eq:dm-rotation}
\end{equation}

The spring displacement vector $\Delta\mathbf l_p^{[i,j]}$ describes the relative displacement between the two vertices connected by ligament $p$ (Supplementary Figure~\ref{fig:discrete-model}(c)). Using the position vectors defined in the Design Space section, the displacement vectors for the four connections are
\begin{equation}
\begin{aligned}
\Delta\mathbf l_1^{[i,j]}={}&
\begin{bmatrix}
a\\
0
\end{bmatrix}
+
\begin{bmatrix}
u^{[i,j+1]}-u^{[i,j]}\\
v^{[i,j+1]}-v^{[i,j]}
\end{bmatrix}
+\mathbf R\left(\theta^{[i,j+1]}\right)\mathbf r_3^{[i,j+1]}
-\mathbf R\left(\theta^{[i,j]}\right)\mathbf r_1^{[i,j]},\\[5pt]
\Delta\mathbf l_2^{[i,j]}={}&
\begin{bmatrix}
0\\
a
\end{bmatrix}
+
\begin{bmatrix}
u^{[i+1,j]}-u^{[i,j]}\\
v^{[i+1,j]}-v^{[i,j]}
\end{bmatrix}
+\mathbf R\left(\theta^{[i+1,j]}\right)\mathbf r_4^{[i+1,j]}
-\mathbf R\left(\theta^{[i,j]}\right)\mathbf r_2^{[i,j]},\\[5pt]
\Delta\mathbf l_3^{[i,j]}={}&
\begin{bmatrix}
-a\\
0
\end{bmatrix}
+
\begin{bmatrix}
u^{[i,j-1]}-u^{[i,j]}\\
v^{[i,j-1]}-v^{[i,j]}
\end{bmatrix}
+\mathbf R\left(\theta^{[i,j-1]}\right)\mathbf r_1^{[i,j-1]}
-\mathbf R\left(\theta^{[i,j]}\right)\mathbf r_3^{[i,j]},\\[5pt]
\Delta\mathbf l_4^{[i,j]}={}&
\begin{bmatrix}
0\\
-a
\end{bmatrix}
+
\begin{bmatrix}
u^{[i-1,j]}-u^{[i,j]}\\
v^{[i-1,j]}-v^{[i,j]}
\end{bmatrix}
+\mathbf R\left(\theta^{[i-1,j]}\right)\mathbf r_2^{[i-1,j]}
-\mathbf R\left(\theta^{[i,j]}\right)\mathbf r_4^{[i,j]}.
\end{aligned}
\label{eq:dm-spring-displacement}
\end{equation}

For quadrilaterals along the boundaries, contributions from missing neighbors are omitted.

\newpage
\subsection{Experimental Calibration}
\label{sec:exp-calibration}
\quad
The discrete model was calibrated against the experimental compression response of an unperturbed $10\times10$ array of squares with diagonal length $a=10~\mathrm{mm}$ (Supplementary Figure~\ref{fig:experimental-calibration}(a)). The calibration considered the initial stiffness, the onset of buckling, and the subsequent post-buckling response, allowing the model to capture both the initial elastic regime and the nonlinear deformation behavior. The parameter set used in the simulations was
\[
\begin{gathered}
k_l=5.5~\mathrm{N/mm},\qquad
k_s=1.65~\mathrm{N/mm},\qquad
k_\theta=1.375~\mathrm{N\,mm/rad},\\
k_{\mathrm{cont}}=5.5~\mathrm{N\,mm/rad},\qquad
\gamma=-0.3.
\end{gathered}
\]

To test whether the calibrated parameters could predict the response of another structure, we fabricated and tested a randomly generated sample. We then simulated its compression response using the same parameters without further adjustment. The predicted stress–strain curve agreed closely with the experimental results, with only modest differences at larger compressive strains (Supplementary Figure~\ref{fig:experimental-calibration} (b)).

\begin{figure}[htbp]
    \centering
    \includegraphics[width=0.9\linewidth]{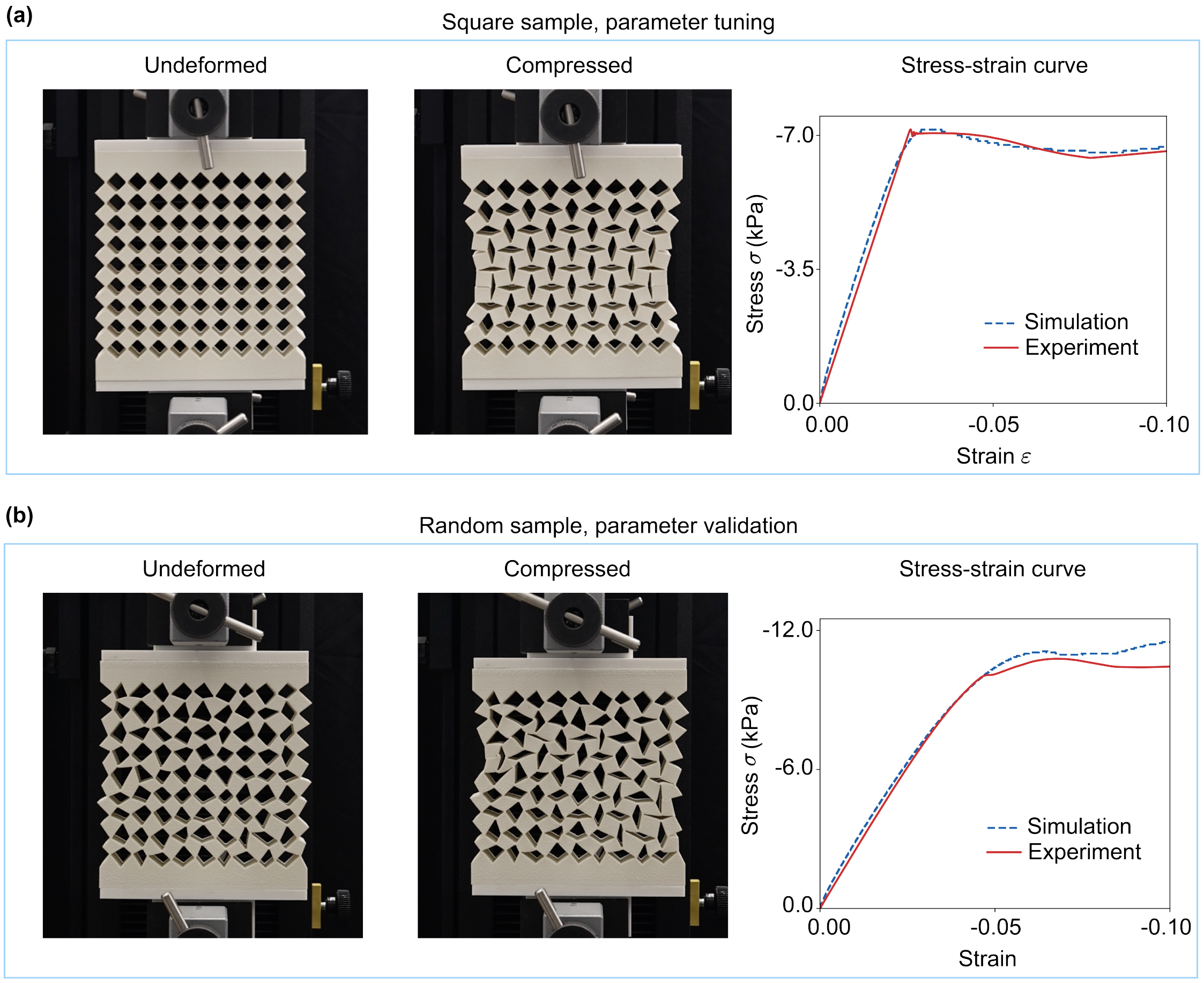}
    \caption{Experimental calibration and validation. (a) Square sample used to calibrate the model. (b) Random sample simulated using the same parameters without further adjustment. Each row shows the undeformed and compressed samples and the experimental (solid red) and simulated (dashed blue) stress--strain curves.}
    \label{fig:experimental-calibration}
\end{figure}

\newpage
\subsection{Linearized Young's Modulus}
\label{sec:linearized-modulus}

\quad
The effective Young's modulus of the metamaterial is calculated from the initial linear response of the discrete model. Using the DOF vector $\mathbf q^{[i,j]}$ defined for each quadrilateral in Eq.~\eqref{eq:dm-dof}, the global DOF vector $\mathbf Q$ is assembled by stacking the vectors of all quadrilaterals row by row as
\begin{equation}
\mathbf Q=
\begin{bmatrix}
\mathbf q^{[1,1]}\\
\vdots\\
\mathbf q^{[1,N_x]}\\
\mathbf q^{[2,1]}\\
\vdots\\
\mathbf q^{[2,N_x]}\\
\vdots\\
\mathbf q^{[N_y,1]}\\
\vdots\\
\mathbf q^{[N_y,N_x]}
\end{bmatrix}
\in\mathbb R^{3N_xN_y}.
\label{eq:dm-global-coordinates}
\end{equation}

Similarly, the internal forces and moments acting on each quadrilateral are assembled in the same order to form the global internal force vector
\begin{equation}
\mathbf f_{\mathrm{int}}(\mathbf Q)=
\begin{bmatrix}
\displaystyle\sum_{p=1}^{4}F_{x,p}^{[1,1]}\\
\displaystyle\sum_{p=1}^{4}F_{y,p}^{[1,1]}\\
\displaystyle\sum_{p=1}^{4}M_p^{[1,1]}\\
\vdots\\
\displaystyle\sum_{p=1}^{4}F_{x,p}^{[N_y,N_x]}\\
\displaystyle\sum_{p=1}^{4}F_{y,p}^{[N_y,N_x]}\\
\displaystyle\sum_{p=1}^{4}M_p^{[N_y,N_x]}
\end{bmatrix}
\in\mathbb R^{3N_xN_y}.
\label{eq:dm-global-internal-force}
\end{equation}

To calculate the initial linear response, we linearize $\mathbf f_{\mathrm{int}}(\mathbf Q)$ using a first-order Taylor expansion about the undeformed state ($\mathbf Q=\mathbf0$):
\begin{equation}
\mathbf f_{\mathrm{int}}(\mathbf Q)\approx\mathbf f_{\mathrm{int}}(\mathbf0)+\left.\frac{\partial\mathbf f_{\mathrm{int}}}{\partial\mathbf Q}\right|_{\mathbf Q=\mathbf0}\mathbf Q.
\label{eq:dm-force-linearization}
\end{equation}

Since the undeformed state is stress-free, $\mathbf f_{\mathrm{int}}(\mathbf0)=\mathbf0$. The initial stiffness matrix is defined as the negative Jacobian of $\mathbf f_{\mathrm{int}}$ with respect to $\mathbf Q$, evaluated at the undeformed state:
\begin{equation}
\mathbf K=-\left.\frac{\partial\mathbf f_{\mathrm{int}}}{\partial\mathbf Q}\right|_{\mathbf Q=\mathbf0},
\label{eq:dm-linear-stiffness}
\end{equation}

With $N_{\mathrm{dof}}=3N_xN_y$, the matrix can be written explicitly as
\begin{equation}
\mathbf K=-\left.
\begin{bmatrix}
\dfrac{\partial f_{\mathrm{int},1}}{\partial Q_1}&\cdots&\dfrac{\partial f_{\mathrm{int},1}}{\partial Q_{N_{\mathrm{dof}}}}\\[6pt]
\vdots&\ddots&\vdots\\[6pt]
\dfrac{\partial f_{\mathrm{int},N_{\mathrm{dof}}}}{\partial Q_1}&\cdots&\dfrac{\partial f_{\mathrm{int},N_{\mathrm{dof}}}}{\partial Q_{N_{\mathrm{dof}}}}
\end{bmatrix}
\right|_{\mathbf Q=\mathbf0}.
\label{eq:dm-stiffness-entries}
\end{equation}

The stiffness matrix $\mathbf K$ is evaluated at the undeformed state ($\mathbf Q=\mathbf0$) and remains constant in the linearized model. Assuming static equilibrium, the external forces and moments balance the internal forces and moments, giving
\begin{equation}
\mathbf f_{\mathrm{ext}}=-\mathbf f_{\mathrm{int}}(\mathbf Q)\approx\mathbf K\mathbf Q,
\label{eq:dm-linear-external-force}
\end{equation}
where $\mathbf f_{\mathrm{ext}}$ is the global vector of external forces and moments.

We partition $\mathbf Q$ into three blocks, $\mathbf Q_b$, $\mathbf Q_f$, and $\mathbf Q_t$, corresponding to the bottom row, the quadrilaterals in the intermediate rows, and the top row, respectively. The stiffness matrix $\mathbf K$ is partitioned accordingly.

The bottom row is fixed, while the top row has zero horizontal displacement and rotation. For a strain $\varepsilon<0$ (compression), the prescribed DOF vector for each top-row quadrilateral is
\begin{equation}
\bar{\mathbf q}^{[N_y,j]}=
\begin{bmatrix}
0\\
\varepsilon(N_y-1)a\\
0
\end{bmatrix},
\qquad j=1,\ldots,N_x.
\label{eq:dm-prescribed-top-unit}
\end{equation}

The prescribed DOF vectors are assembled into $\bar{\mathbf Q}_t$ as
\begin{equation}
\bar{\mathbf Q}_t=
\begin{bmatrix}
\bar{\mathbf q}^{[N_y,1]}\\
\vdots\\
\bar{\mathbf q}^{[N_y,N_x]}
\end{bmatrix}
\in\mathbb R^{3N_x}.
\label{eq:dm-prescribed-top}
\end{equation}

Since no external forces or moments act directly on the quadrilaterals in the intermediate rows, the global external force vector $\mathbf f_{\mathrm{ext}}$ contains only the forces and moments exerted by the bottom and top boundary constraints, denoted by $\mathbf f_b$ and $\mathbf f_t$, respectively:
\begin{equation}
\mathbf f_{\mathrm{ext}}=
\begin{bmatrix}
\mathbf f_b\\
\mathbf0\\
\mathbf f_t
\end{bmatrix}.
\label{eq:dm-external-force-vector}
\end{equation}

With $\mathbf Q_b=\mathbf0$ and $\mathbf Q_t=\bar{\mathbf Q}_t$, the global equilibrium equation (Eq.~\eqref{eq:dm-linear-external-force}) becomes
\begin{equation}
\begin{bmatrix}
\mathbf K_{bb}&\mathbf K_{bf}&\mathbf K_{bt}\\
\mathbf K_{fb}&\mathbf K_{ff}&\mathbf K_{ft}\\
\mathbf K_{tb}&\mathbf K_{tf}&\mathbf K_{tt}
\end{bmatrix}
\begin{bmatrix}
\mathbf0\\
\mathbf Q_f\\
\bar{\mathbf Q}_t
\end{bmatrix}
=
\begin{bmatrix}
\mathbf f_b\\
\mathbf0\\
\mathbf f_t
\end{bmatrix}.
\label{eq:dm-block-equilibrium}
\end{equation}

Expanding the three block rows gives
\begin{equation}
\begin{aligned}
\mathbf f_b&=\mathbf K_{bf}\mathbf Q_f+\mathbf K_{bt}\bar{\mathbf Q}_t,\\
\mathbf K_{ff}\mathbf Q_f&=-\mathbf K_{ft}\bar{\mathbf Q}_t,\\
\mathbf f_t&=\mathbf K_{tf}\mathbf Q_f+\mathbf K_{tt}\bar{\mathbf Q}_t.
\end{aligned}
\label{eq:dm-block-equations}
\end{equation}

Solving Eq.~\eqref{eq:dm-block-equations} gives
\begin{equation}
\begin{aligned}
\mathbf f_b&=\left(\mathbf K_{bt}-\mathbf K_{bf}\mathbf K_{ff}^{-1}\mathbf K_{ft}\right)\bar{\mathbf Q}_t,\\
\mathbf f_t&=\left(\mathbf K_{tt}-\mathbf K_{tf}\mathbf K_{ff}^{-1}\mathbf K_{ft}\right)\bar{\mathbf Q}_t.
\end{aligned}
\label{eq:dm-boundary-reactions}
\end{equation}

The forces and moments exerted by the top boundary constraint are collected in
\begin{equation}
\mathbf f_t=
\begin{bmatrix}
f_u^{[N_y,1]}\\
f_v^{[N_y,1]}\\
f_\theta^{[N_y,1]}\\
\vdots\\
f_u^{[N_y,N_x]}\\
f_v^{[N_y,N_x]}\\
f_\theta^{[N_y,N_x]}
\end{bmatrix}
\in\mathbb R^{3N_x},
\label{eq:dm-top-reaction-vector}
\end{equation}
where $f_u^{[N_y,j]}$, $f_v^{[N_y,j]}$, and $f_\theta^{[N_y,j]}$ are the horizontal force, vertical force, and moment acting on quadrilateral $[N_y,j]$, respectively.

The total vertical force on the top row is
\begin{equation}
F_t=\sum_{j=1}^{N_x}f_v^{[N_y,j]}.
\label{eq:dm-top-force}
\end{equation}

The corresponding nominal stress is
\begin{equation}
\sigma=\frac{F_t}{A}=\frac{1}{A}\sum_{j=1}^{N_x}f_v^{[N_y,j]},
\label{eq:dm-linear-stress}
\end{equation}
where $A$ is the sample loading area. Both $F_t$ and $\sigma$ are negative under compression.

Within the linearized model, the stress is proportional to the prescribed strain. The effective Young's modulus is therefore
\begin{equation}
E=\frac{\sigma}{\varepsilon}=\frac{1}{A\varepsilon}\sum_{j=1}^{N_x}f_v^{[N_y,j]}.
\label{eq:dm-linear-modulus}
\end{equation}

\section{ML Training Details}

\label{sec:exp-appendix}

\paragraph{Dataset}
The dataset was generated by partial inverse design of the Young modulus (targets from minimization to maximization, at most 50 iterations per run) with CCSA, Adam with augmented Lagrangian and SLSQP, giving 134 designs. Mirror symmetries, uniform translations and alternating perturbations, alone or combined (up to three at a time), that satisfy the displacement constraint expanded them to $31{,}638$ samples ($\times 235$). We split the data 95/5 by source design, so that no variant of a validation design is seen in training.

\paragraph{Encoder and latent posterior}
The patch tokens of the ViT-base encoder are aggregated by a learned attention pooling into a single 768-dimensional feature vector. Two linear layers then map this vector to the mean $\boldsymbol{\mu} \in \mathbb{R}^{8}$ and the log-variance $\log \boldsymbol{\sigma}^2 \in \mathbb{R}^{8}$ of the latent posterior.

\paragraph{Decoder}
The latent vector $\mathbf{z} \sim \mathcal{N}(\boldsymbol{\mu}, \mathrm{diag}(\boldsymbol{\sigma}^2))$, sampled with the reparametrization trick, is passed through two independent multi-layer perceptrons (MLPs), one for the horizontal and one for the vertical parameters.

\paragraph{Training}
The ViT-VAE is trained for 20 epochs with a cosine learning rate schedule, from a maximum rate of $10^{-4}$ to a minimum of $5\times10^{-6}$ with batches of $64$ image-parameters pairs. The KL-divergence weight $\beta$ increases linearly from $10^{-4}$ to $10^{-2}$ over the full training.

\paragraph{Transfer Analysis}
We fit the PCA using all the designs we found during optimization runs of that specific property. PCA for the energy experiment will be fitted using all energy designs found by TRAMA, CCSA or CMA-ES for the energy efficiency problem.

\section{Optimization Details}
\label{sec:opt-appendix}

\paragraph{Energy Efficiency Maximization}
We swept the stopping tolerance of CCSA and the maximum number of evaluations of CMA-ES.
Tables~\ref{tab:si-ccsa-tol} and~\ref{tab:si-cmaes-budget} show the effect of these parameters on
energy-efficiency maximization.

\begin{table}[h]
\centering
\caption{CCSA on the energy-efficiency task for different stopping tolerances (relative change of the
objective). Evaluations count forward and gradient evaluations. Values are mean $\pm$ standard deviation over
5 seeds. The configuration used in the main text is marked with an asterisk.}
\label{tab:si-ccsa-tol}
\begin{tabular}{ccc}
\hline
Tolerance & Energy efficiency (\%) & Evaluations \\
\hline
$10^{-3}$        & $82.1 \pm 3.5$ & $171 \pm 87$ \\
$10^{-4}$        & $84.3 \pm 3.0$ & $432 \pm 86$ \\
$10^{-5}$      & $84.6 \pm 2.6$ & $697 \pm 35$ \\
$10^{-6}$$^{*}$  & $85.3 \pm 2.0$ & $631 \pm 222$ \\
\hline
\end{tabular}
\end{table}

\begin{table}[h]
\centering
\caption{CMA-ES on the energy-efficiency task for different evaluation budgets. Values are mean $\pm$
standard deviation over five seeds; all runs use the full budget. The configuration used in the main text is
marked with an asterisk.}
\label{tab:si-cmaes-budget}
\begin{tabular}{cc}
\hline
Evaluation budget & Energy efficiency (\%) \\
\hline
900        & $82.1 \pm 0.6$ \\
1200 & $82.5 \pm 0.8$ \\
1500       & $83.0 \pm 0.4$ \\
2000       & $83.3 \pm 0.5$ \\
3000       & $84.6 \pm 0.7$ \\
8000$^{*}$       & $86.1 \pm 0.7$ \\
\hline
\end{tabular}
\end{table}

We ultimately chose a tolerance of $10^{-6}$ for CCSA and $n_{\text{eval}} = 1200$ for CMA-ES.
TRAMA was run with $\lambda = 10^{-4}$.

\paragraph{Stress-strain curve matching}

TRAMA was run with $\lambda = 10^{-4}$ and a tolerance of $10^{-4}$. CMA-ES was run with a maximum number of iterations equal to $1500$. CCSA was run with a tolerance of $10^{-4}$.

\fi

\end{document}